\documentclass{article}

\usepackage{arxiv}

\usepackage[utf8]{inputenc} 
\usepackage[T1]{fontenc}    
\usepackage{hyperref}       
\usepackage{url}            
\usepackage{booktabs}       
\usepackage{amsfonts}       
\usepackage{nicefrac}       
\usepackage{microtype}      
\usepackage{lipsum}
\usepackage{graphicx}
\graphicspath{ {./images/} }

\RequirePackage{amsthm,amsmath,amsfonts,amssymb}
\RequirePackage[numbers]{natbib}

\usepackage{algorithm,appendix}
\usepackage{algorithmic,multirow,booktabs,makecell}

\newtheorem{theorem}{Theorem}
\newtheorem{corollary}{Corollary}
\newtheorem{proposition}{Proposition}
\newtheorem{lemma}{Lemma}

\title{A penalized online sequential test of heterogeneous treatment effects for generalized linear models}

\author{
 Zhiqing Fang \\
  School of Statistics and Management, \\
  Shanghai University of Finance and Economics
  \texttt{fang.zhiqing@163.sufe.edu.cn} \\
   \And
 Shuyan Chen \\
  School of Management, \\
  University of Science and Technology of China
  \texttt{chenst@ustc.edu.cn} \\
  \And
 Xin Liu \\
  School of Statistics and Management, \\
  Shanghai University of Finance and Economics
  \texttt{liu.xin@mail.shufe.edu.cn} \\
}

\begin{document}

\maketitle

\begin{abstract}
Identification of heterogeneous treatment effects (HTEs) has been increasingly popular and critical in various penalized strategy decisions using the A/B testing approach, especially in the scenario of a consecutive online collection of samples. However, in high-dimensional settings, such an identification remains challenging in the sense of lack of detection power of HTEs with insufficient sample instances for each batch sequentially collected online. In this article, a novel high-dimensional test is proposed, named as the penalized online sequential test (POST), to identify HTEs and select useful covariates simultaneously under continuous monitoring in generalized linear models (GLMs), which achieves high detection power and controls the Type I error. A penalized score test statistic is developed along with an extended $p$-value process for the online collection of samples, and the proposed POST method is further extended to multiple online testing scenarios, where both high true positive rates and under-controlled false discovery rates are achieved simultaneously. Asymptotic results are established and justified to guarantee properties of the POST, and its performance is evaluated through simulations and analysis of real data, compared with the state-of-the-art online test methods. Our findings indicate that the POST method exhibits selection consistency and superb detection power of HTEs as well as excellent control over the Type I error, which endows our method with the capability for timely and efficient inference for online A/B testing in high-dimensional GLMs framework.
\end{abstract}


\section{Introduction}
As an increasingly popular randomized control experiment method, the A/B testing approach has been widely adopted in various industries for iterative enhancement of products and technologies \cite{kohavi2009controlled,shi_2023,yang_2017}. 
The primary goal of an A/B testing is to assess whether a statistically significant improvement exists in the treatment group over the control by manipulating different covariates. Usually, such an improvement is quantified by the average treatment effect (ATE) between the two groups. When the treatment effect is found to vary among individuals either in magnitude or direction, the heterogeneous treatment effect (HTE) may be preferred, and detecting HTE among individuals can help identify subgroups and assist formulation of personalized strategies for individuals \cite{nie_2021,sun_2021,wager_2018estimation}. From a statistical perspective, hypothesis tests are usually adopted to detect nullity of treatment effects in A/B testing, and hence such a scheme may be vulnerable to the Type I and Type II errors. On one hand, the Type I error is akin to adopting an erroneous strategy when HTE exists, while on the other hand, the Type II error may lead to a missing correct decision and adopting an overly conservative stance. To control both of them in practice, a large sample size is usually required to obtain the minimum detectable effect (MDE), which is proved to be a highly effective strategy \cite{lai1979nonlinear,lai1997optimal}. 

Nevertheless, in practice, samples of individuals may be collected in different batches in a sequential manner, sometimes referred to as online collection, and each batch may only contain a limited number of instances rather than an excessively large one owing to the increasing opportunity costs \cite{diagne_2021,stieglitz_2018}. Consequently, it is crucial to establish an effective A/B testing method that is suitable for online sequentially-collected samples of limited sizes while attaining sufficient statistical power \cite{lehmann1986testing,liu_2021,zhou_2023}. To deal with the problem, one potential approach is to employ the so-called sequential test (ST) with a sequence of probability ratios being its test statistic \cite{wald1992sequential}. Such an online procedure enables intermediate analysis and decision-making in a sequential manner, and has gained popularity in online A/B testing due to its flexibility in continuously monitoring the experiment and stopping at a data-driven termination, while controlling the Type I error \citep{Balsubramani2015, johari2017peeking, Johari2015}. As soon as a significant treatment effect is detected, the whole online experiment will be terminated immediately. A few variants of the ST method are developed to detect ATE and proved to be effective in low dimensions, including the mixture sequential probability ratio test (mSPRT \cite{robbins1970statistical}) and maximized sequential probability ratio test (MaxSPRT \cite{kulldorff2011maximized}), where the mSPRT is incorporated into A/B testing \cite{johari2017peeking, johari2015always} owing to its high statistical power \cite{robbins1974expected} and its nearly optimality in expected stopping time \cite{pollak1978optimality}, so that the continuous control over the Type I error is achieved.
    
Nonetheless, several challenging issues remain unsolved. One is that the ST method and its variants may not be suitable for detecting HTE, since its test statistic using the likelihood ratio lacks of an explicit form with a conjugate prior probability, and the Type I error may not be controlled anymore. A sequential score test (SST \citep{yu2020new}) is developed, where the test statistic is an integration of the ratio of score functions with respect to the conjugate prior probability, while it may be less accurate when the sample size is insufficient, especially for the online collection scenarios. Another issue is that, only a part of variables are expected to be useful for detecting HTE with limited sample sizes, even though there may be large numbers of covariates for manipulation \cite{meier_2008,tibshirani1996regression,zhu_2020,zou2006adaptive}, while the current ST method may not be able to capture the potentially useful ones, and it is imperative to introduce variable selection methods into the online HTE detection procedure. One typical way to control the dimension is to employ the penalized method, and several penalty functions are available for penalized likelihood estimation with excellent accuracy and selection consistency, including the Least Absolute Shrinkage and Selection Operator (Lasso \cite{tibshirani1996regression}), the Adaptive Lasso (AdaLasso \cite{zou2006adaptive}), the Smoothly Clipped Absolute Deviation (SCAD \cite{fan2001variable}) and Minimax Concave Penalty (MCP \cite{zhang2010nearly}), among others. 
A few studies attempt to detect high-dimensional treatment effects. For example, a penalized regression is employed for estimating marginal effects and treatment propensities, and a quasi-oracle property is demonstrated \cite{nie_2021}. Further, a general theory of regression adjustment on two Lasso-adjusted treatment effect estimators is developed upon the adjusted ordinary least square estimator under covariate-adaptive randomization, and the two estimators are proved to be optimal in their own classes, respectively \cite{liu2023lasso}. Also, a finite-sample-unbiased treatment effect estimator is obtained using the cross-estimation method, harnessing high-dimensional regression adjustments \cite{wager2016high}. To the best of our knowledge, very limited research has focused on online detection of HTEs in a high-dimensional setting and simultaneous selection of useful covariates with limited sample sizes, and hence research gap still remains. 

Inspired by these statistical challenges, a novel penalized online sequential test (POST) method is proposed in this article, which facilitates an online exploration of high-dimensional heterogeneous treatment effects in a sequential manner with limited sample sizes, and identifies possibly useful covariates that contribute to HTEs under the generalized linear models framework. A test statistic is constructed based on the likelihood ratio of score functions with unknown parameters being estimated by the penalized maximum likelihood estimation, and the asymptotic distributions of the test statistic are obtained under both the null and alternative hypotheses, respectively. Further, an extended $p$-value process is proposed for the POST method with online collection of samples, and the proposed POST method is further extended to multiple online testing scenarios, where both high true positive rates and under-controlled false discovery rates are achieved simultaneously. The proposed method is demonstrated to exhibit variable selection consistency and an asymptotically close-to-one detection power of HTE, effectively controlling the Type I error at any given batch of online collected samples. Numerical investigations on both simulated and real datasets have supported the superb performance of the proposed POST method in different scenarios.

The rest of the article is organized as follows. In Section 2, the proposed POST method is introduced and the test procedure is outlined, along with theoretical justifications. The performance of the proposed POST method is examined by simulation studies in Section 3 and real data analysis in Section 4. In Section 5, we conclude the article, followed by a discussion.

\section{Methodology}
\subsection{Treatment effects and the sequential test}
To start with, consider an i.i.d. sample $\{\left(Y_i, \mathbf{X}_i, A_i\right)\}_{i=1}^n$, where $Y_i$ and $\mathbf{X}_i$ denote the response and the $(p+1)$-dimensional covariate vector including an intercept for the $i$-th subject, and $A_i$ is a binary indicator, taking 1 if the subject gets treatment and 0 otherwise, respectively. Usually, the treatment effect is quantified as the difference between the responses with and without treatment for the $i$-th subject, and the average treatment effect (ATE) is employed as a popular measurement, defined as $E(Y|A_i=1, \mathbf{X})-E(Y|A_i=0, \mathbf{X})$. Owing to its popularity in personalized strategies for subgroups, we focus on HTE detection under the framework of generalized linear models in this article. Specifically, assume $Y_i$ is randomly drawn from a certain distribution in the exponential family, whose density function is described by
\begin{equation}\label{eq:GLM}
        f_{Y_i}\left(y_i \mid \gamma_i, \phi\right)=\exp \left\{\frac{y_i \gamma_i-b\left(\gamma_i\right)}{a_i(\phi)}+c\left(y_i, \phi\right)\right\},\quad i=1,\ldots,n,
    \end{equation}
where $a_i(\cdot), b(\cdot)$ and $c(\cdot, \cdot)$ are known functions, $\gamma_i$ is the canonical parameter, and $\phi$ is a typically known dispersion parameter, and the mean $\mu_i$ of $Y_i$ and its variance $\operatorname{Var}\left(Y_i\right)$ are obtained by
\begin{equation*}
\mu_i=\mathbb{E}\left(Y_i\right)=b^{\prime}\left(\gamma_i\right), \quad \operatorname{Var}\left(Y_i\right)=a_i(\phi) \cdot b^{\prime \prime}\left(\gamma_i\right) .
\end{equation*}
Usually, a linear function $g(\cdot)$ is assumed to link $\mu_i$ with the covariates $\mathbf{X}_i$ and the treatment indicator $A_i$ as
\begin{equation}\label{eq:ATE}
g\left(\mu_i\right)=\eta_i=\boldsymbol{\theta}^\top \mathbf{X}_i+\beta A_i,
\end{equation}
where $\beta$ and $\boldsymbol{\theta}$ indicate the ATE and the baseline main effects, respectively. Consequently, to detect whether a treatment effect exists or not relies on statistical inference of $\beta$. Usually in GLMs, the maximum likelihood estimation is employed to obtain the estimate of $\beta$ along with $\boldsymbol{\theta}$, and a hypothesis test of nullity on $\beta$ is typically considered, i.e. $H_0: \beta = \beta_0 \quad \mbox{v.s.} \quad H_1: \beta = \beta_0 + \frac{\delta}{\sqrt{n}} (\delta \neq 0)$, so that the Type I error is controlled and the power can be analyzed by changing $\delta$ given a sample size $n$. 

Specifically for online experiments, a sequential test is further considered, which takes sequential probability ratio as its statistic, $T = \frac{p_{1 n}\left(y_1, \ldots, y_n\right)}{p_{0 n}\left(y_1, \ldots, y_n\right)},$ where $p_{k n}\left(y_1, \ldots, y_n\right)$ denotes the probability density in the $n$-dimensional sample space calculated under $H_k$ for $k=0$ or 1. Then, given $\alpha$ and $\omega$ as the pre-specified upper-bounds of the Type I and II errors, $A=\frac{1-\omega}{\alpha}$ and $B=\frac{\omega}{1-\alpha}$, one of the following three decisions will be made: (i) accept $H_0$ if $T \leq B$, (ii) reject $H_0$ if $T \geq A$, (iii) take more observations if $B<T<A$. In general, the decision rule for sequential testing may be described by a pair $(T(\alpha), d(\alpha))$ parameterized by a bounded Type I error rate $\alpha\in (0,1)$, where $T(\alpha)$ is a stopping time, usually indicated by the cumulative sample size in online collection, and $d(\alpha)$ is a binary indicator for rejection, taking 1 if the null is rejected. Usually, $T(\alpha)$ is non-increasing in $\alpha$ and $d(\alpha)$ is non-decreasing. For a given sequential test $(T(\alpha), d(\alpha))$, the ATE can be detected by the so-called \textit{always valid $p$-value process} $(p_n)^{\infty}_{n=1}$, where
\[
p_n=\inf\{\alpha:T(\alpha)\leq n, d(\alpha)=1\},
\]
and the decision of a sequential test is obtained as 
\[
T(\alpha)=\inf\{n:p_n\leq\alpha\}, \quad d(\alpha)=1\{T(\alpha)<\infty\}.
\]

\subsection{Heterogeneous treatment effects and the penalized online sequential test}
To identify HTEs in a high-dimensional setting with a sequence of samples which are consecutively collected online, we propose a penalized online sequential test (POST), where a penalized maximum likelihood estimator is adopted to estimate unknown parameters in GLMs first of all, and then a sequential score test is created to identify the potential HTE. Accordingly, an interaction term is proposed to be included into \eqref{eq:ATE}
\begin{equation}\label{eq:HTE}
g\left(\mu_i\right)=\eta_i=\boldsymbol{\theta}^\top \mathbf{X}_i+\left(\boldsymbol{\beta}^\top \mathbf{X}_i\right) A_i, 
\end{equation}
so that $\boldsymbol{\beta}$ now denotes the HTEs, since the difference in $\eta_i$ between the treatment group and the control is easily obtained as $\boldsymbol{\beta}^\top\mathbf{X}_i$ for the $i$-th subject. Similarly, to examine whether HTEs exist in \eqref{eq:HTE}, a hypothesis test of nullity on $\boldsymbol{\beta}$ is considered. Specifically, we consider a test with a local alternative as 
\begin{equation}\label{eq:HTEtest}
   H_0: \boldsymbol{\beta} = \boldsymbol{\beta}_0 \qquad \mbox{v.s.} \qquad
   H_1: \boldsymbol{\beta} = \boldsymbol{\beta}_0 + \frac{\boldsymbol{\delta}}{\sqrt{n}} \quad(\boldsymbol{\delta} \neq 0),
\end{equation}
where $\boldsymbol{\delta}$ is prespecified in the local alternatives to analyze the statistical power. To construct the test statistic, we start with the likelihood function for a certain online collected sample $\{\left(Y_i, \mathbf{X}_i, A_i\right)\}_{i=1}^n$ which contains $n$ observations, which is described by
\begin{equation}\label{eq:likelihood}
    L(\boldsymbol{\theta}, \boldsymbol{\beta}):=\exp \left\{\sum_{i=1}^{n}{\frac{Y_i \gamma_i-b\left(\gamma_i\right)}{a_i(\phi)}} + \sum_{i=1}^{n}{c\left(Y  _i, \phi\right)}\right\}.
\end{equation} where the mean of $Y_i$ is associated with $\mathbf{X}_i$ and $A_i$ by \eqref{eq:HTE}. Let $\mathbf{S}_{n, \boldsymbol{\beta}}^{(1)}\left(\boldsymbol{\theta}, \boldsymbol{\beta}_0\right)$ be the score function for the treatment group $(A_i=1)$ under $H_0$
\begin{equation}\label{eq:scoreH0}
    \mathbf{S}_{n, \boldsymbol{\beta}}^{(1)}\left(\boldsymbol{\theta}, \boldsymbol{\beta}_0\right):=\left.\sum_{i=1}^n\left(\frac{\partial \mu_i^{(1)}(\boldsymbol{\beta}, \boldsymbol{\theta})}{\partial \boldsymbol{\beta}^\top} \cdot \frac{\left(Y_i^{(1)}-\mu_i^{(1)}(\boldsymbol{\beta}, \boldsymbol{\theta})\right)}{a_i(\phi) \cdot V_i^{(1)}(\boldsymbol{\beta}, \boldsymbol{\theta})}\right)\right|_{\boldsymbol{\beta}=\boldsymbol{\beta}_0},
\end{equation}
where $\mu_i^{(1)}(\boldsymbol{\beta}, \boldsymbol{\theta})$=$\mathbb{E}\left(Y_i \mid A_i=1, \mathbf{X}_i\right)$, $a_i(\phi) \cdot V_i^{(1)}(\boldsymbol{\beta}, \boldsymbol{\theta})$=$\operatorname{Var}\left(Y_i \mid A_i=1, \mathbf{X}_i\right)$ and assume $a_i(\phi)=a(\phi)$ for all $i$, $a(\cdot)$ is known. Further, denote $\overline{\mathbf{S}}_n$ as the estimated average score for the treatment group  under $H_0$
\begin{equation}\label{eq:average score}
        \overline{\mathbf{S}}_n:=\frac{1}{n} \mathbf{S}_{n, \boldsymbol{\beta}}^{(1)}\left(\hat{\boldsymbol{\theta}}, \boldsymbol{\beta}\right)|_{\boldsymbol{\beta}=\boldsymbol{\beta}_0},
\end{equation}
where $\hat{\boldsymbol{\theta}}$ is the estimated baseline main effects calculated with the data from the control group ($A_i=0$).
Consequently, the test statistic $\tilde{\Lambda}_n$ is proposed in the form of a likelihood ratio as
\begin{equation}\label{eq:llratio}
\tilde{\Lambda}_n=2\log\frac{\underset{\beta \in H_1}{\max}\quad\psi_{(\overline{\mathbf{I}}_n^{(1)}(\boldsymbol{\beta_0})\cdot(\boldsymbol{\beta}-\boldsymbol{\beta}_0), \frac{\overline{\mathbf{v}}_n(\hat{\boldsymbol{\theta}})}{n})}(\overline{\mathbf{S}}_n)}{\psi_{(\mathbf{0}, \frac{\overline{\mathbf{v}}_n(\hat{\boldsymbol{\theta}})}{n})}(\overline{\mathbf{S}}_n)},
\end{equation}
where
\begin{itemize}
\item $\psi_{(\boldsymbol{\mu}, \boldsymbol{\Sigma})}(\cdot)$ denotes the probability density function of a multivariate normal distribution with the mean vector $\boldsymbol{\mu}$ and covariance matrix $\boldsymbol{\Sigma}$,

\item $\overline{\mathbf{I}}_n^{(1)}(\boldsymbol{\beta}) = -\frac{1}{n} \frac{\partial \mathbf{S}_{n, \boldsymbol{\beta}}^{(1)}\left(\boldsymbol{\theta}, \boldsymbol{\beta}\right)}{\partial \boldsymbol{\beta}}$,

\item $\overline{\mathbf{V}}_n(\boldsymbol{\theta})=\overline{\mathbf{I}}_n^{(1)}(\boldsymbol{\theta})+\overline{\mathbf{I}}_n^{(1)}(\boldsymbol{\theta})\boldsymbol{\hat{\Sigma}} (\boldsymbol{\theta})\overline{\mathbf{I}}_n^{(1)}(\boldsymbol{\theta})$,

\item $\overline{\mathbf{I}}_n^{(1)}(\boldsymbol{\theta}) = \left.-\frac{1}{n} \frac{\partial \mathbf{S}_{n, \boldsymbol{\beta}}^{(1)}\left(\boldsymbol{\theta}, \boldsymbol{\beta}\right)}{\partial \boldsymbol{\theta}}\right|_{\boldsymbol{\beta}=\boldsymbol{\beta}_0} $,

\item $\boldsymbol{\hat{\Sigma}}(\boldsymbol{\theta})$ is the estimator of covariance matrix of $\boldsymbol{\hat{\theta}}$ calculated with data from the control group $(A_i=0)$.
\end{itemize}

However, when obtain the $\hat{\boldsymbol{\theta}}$ in \eqref{eq:average score} in high-dimensional settings, the standard maximum likelihood estimation may fail to identify the useful covariates, and hence the HTE parameter $\boldsymbol{\beta}$ may be tested inaccurately, especially with limited sizes of samples in online detection procedure. Accordingly, the penalized likelihood estimate is proposed to identify the useful covariates in our case. Explicitly, instead of minimizing $-L(\boldsymbol{\theta}, \boldsymbol{\beta})$ in \eqref{eq:likelihood} directly, we propose minimizing its penalized version with an objective function with respect to $\boldsymbol{\theta}$,
\begin{equation*} 
    L^{*}(\boldsymbol{\theta}, \boldsymbol{\beta}) = - L(\boldsymbol{\theta}, \boldsymbol{\beta}) + J_{\lambda}(\boldsymbol{\theta}), 
\end{equation*}
where $J_{\lambda}(\boldsymbol{\theta})$ is a scalar penalty function that penalizes $\boldsymbol{\theta}$ with a tuning parameter $\lambda$ that controls the sparsity of $\boldsymbol{\theta}$ and further helps identify the useful covariates. Usually, a penalty function $J_{\lambda}(\cdot)$ has two properties:
\begin{itemize}
    \item being positive and differentiable in the support,
    \item using non-negative tuning parameter(s) to control its range.
\end{itemize}
Note that either convex or non-convex penalties may be employed. Specifically, we consider different forms of penalty functions for $J_{\lambda}(\cdot)$, including
\begin{itemize}
    \item AdaLasso: $J_{\lambda}(\boldsymbol{\theta})=\sum_{j=1}^{p+1}J_{\lambda}(\theta_j)$ where $$J_{\lambda}(\theta_j) = w_j\left|\theta_j\right|$$
    with $w_j=\left|\tilde{\theta}_j\right|^{-1}$ and $\tilde{\theta}_j$ is an initial estimate of ${\theta_j}$ such as the coefficient of ridge regression,
    \item SCAD: $J_{\lambda}(\boldsymbol{\theta})=
    \sum_{j=1}^{p+1}J_{\lambda}(\theta_j \mid \gamma)$ where $$J_{\lambda}(\theta_j \mid \gamma) = 
    \begin{cases} \lambda|\theta_j| & \text { if }|\theta_j| \leq \lambda, \\ \frac{2 \gamma \lambda|\theta_j|-\theta_j^2-\lambda^2}{2(\gamma-1)} & \text { if } \lambda<|\theta_j|<\gamma \lambda, \\ \frac{\lambda^2(\gamma+1)}{2} & \text { if }|\theta_j| \geq \gamma \lambda\end{cases}$$ where $\gamma>2$ controlling the concavity of the penalty,
    \item MCP: $J_{\lambda}(\boldsymbol{\theta})=
    \sum_{j=1}^{p+1}J_{\lambda}(\theta_j \mid \gamma)$ where $$J_{\lambda}(\theta_j \mid \gamma)= 
    \begin{cases}\lambda|\theta_j|-\frac{\theta_j^2}{2 \gamma} & \text { if }|\theta_j| \leq \gamma \lambda, \\ \frac{1}{2} \gamma \lambda^2 & \text { if }|\theta_j|>\gamma \lambda\end{cases}
    $$ where $\gamma>1$ controlling the concavity of the penalty.
\end{itemize}

\noindent With the penalized estimator $\hat{\boldsymbol{\theta}}$, all components of the test statistic $\tilde{\Lambda}_n$ in \eqref{eq:llratio} are obtained. Now we derive the asymptotic distribution of $\overline{\mathbf{S}}_n$ under $H_0$ and $H_1$ in \eqref{eq:HTEtest} and further that of $\tilde{\Lambda}_n$.

\begin{theorem}\label{theo: distribution of average score}
    For a generalized linear model in \eqref{eq:GLM}, a link function $\eta_i$ in \eqref{eq:HTE} and $\overline{\mathbf{S}}_n$ in \eqref{eq:average score}, define 
    \begin{eqnarray*}
        \mathbf{I}^{(1)}(\boldsymbol{\beta}):=\mathbb{E}_{(\mathbf{X}, \mathbf{Y})}\left[\overline{\mathbf{I}}_n^{(1)}(\boldsymbol{\beta})\right]=\mathbb{E}_{(\mathbf{X}, \mathbf{Y})}\left[-\frac{1}{n} \frac{\partial \mathbf{S}_{n, \boldsymbol{\beta}}^{(1)}\left(\boldsymbol{\theta}, \boldsymbol{\beta}\right)}{\partial \boldsymbol{\beta}}\right], \quad and \\
        \mathbf{I}^{(1)}(\boldsymbol{\theta}):=\mathbb{E}_{(\mathbf{X}, \mathbf{Y})}\left[\overline{\mathbf{I}}_n^{(1)}(\boldsymbol{\theta})\right]=\left.\mathbb{E}_{(\mathbf{X}, \mathbf{Y})}\left[-\frac{1}{n} \frac{\partial \mathbf{S}_{n, \boldsymbol{\beta}}^{(1)}\left(\boldsymbol{\theta}, \boldsymbol{\beta}\right)}{\partial \boldsymbol{\theta}}\right]\right|_{\boldsymbol{\beta}=\boldsymbol{\beta}_0}.
    \end{eqnarray*}
    Then, under $H_0: \boldsymbol{\beta}=\boldsymbol{\beta}_0$,
    \begin{equation*}
        \sqrt{n}~\overline{\mathbf{S}}_n {\stackrel{d}{\longrightarrow}} \mathbf{M V N}_{p+1}\left(\mathbf{0}, \mathbf{V}\left(\boldsymbol{\theta}_0\right)\right)
    \end{equation*}
    whereas under $H_1: \boldsymbol{\beta}=\boldsymbol{\beta}_0+\frac{\boldsymbol{\delta}}{\sqrt{n}}$, 
    \begin{equation*}
        \sqrt{n}\left(\overline{\mathbf{S}}_n-\mathbf{I}^{(1)}\left(\boldsymbol{\beta}_0\right)\left(\boldsymbol{\beta}-\boldsymbol{\beta}_0\right)\right) {\stackrel{d}{\longrightarrow}} \mathbf{M V N}_{p+1}\left(\mathbf{0}, \mathbf{V}\left(\boldsymbol{\theta}_0\right)\right)
    \end{equation*}
    where $\mathbf{V}(\boldsymbol{\theta})=\mathbf{I}^{(1)}(\boldsymbol{\theta})+\mathbf{I}^{(1)}(\boldsymbol{\theta})\boldsymbol{\Sigma}(\boldsymbol{\theta}) \mathbf{I}^{(1)}(\boldsymbol{\theta})$, $\boldsymbol{\Sigma(\boldsymbol{\theta})}$ is the covariance matrix of the penalized estimator of $\boldsymbol{\theta}$, and $\boldsymbol{\theta_0}$ is the true value of the nuisance parameter.
\end{theorem}

\begin{theorem}\label{theo:distribution of LLR}
    For a generalized linear model in \eqref{eq:GLM}, a link function $\eta_i$ in \eqref{eq:HTE}, $\tilde{\Lambda}_n$ in \eqref{eq:llratio} and $\mathbf{I}^{(1)}(\boldsymbol{\beta})$, $\mathbf{I}^{(1)}(\boldsymbol{\theta})$ in Theorem~\ref{theo: distribution of average score}, under $H_0:\boldsymbol{\beta}=\boldsymbol{\beta}_0$,
    \begin{equation*}
        \tilde{\Lambda}_n {\stackrel{d}{\longrightarrow}} \chi^2_{p+1}\left(u_0\right),
    \end{equation*} where $u_0=-\boldsymbol{\beta}_0^\top{\mathbf{I}^{(1)}(\boldsymbol{\beta}_0)}\left(\frac{\mathbf{V}(\boldsymbol{\theta}_0)}{n}\right)^{-1}{\mathbf{I}^{(1)}(\boldsymbol{\beta}_0)}\boldsymbol{\beta}_0$ is the non-centrality parameter, and $p+1$ is the degree of freedom,
    whereas under $H_1: \boldsymbol{\beta}=\boldsymbol{\beta}_0+\frac{\boldsymbol{\delta}}{\sqrt{n}}$, 
    \begin{equation*}
        \tilde{\Lambda}_n {\stackrel{d}{\longrightarrow}} \chi^2_{p+1}\left(u_1\right),
    \end{equation*} where $u_1=(\frac{\boldsymbol{\delta}}{\sqrt{n}}-\boldsymbol{\beta}_0)^\top{\mathbf{I}^{(1)}(\boldsymbol{\beta}_0)}\left(\frac{\mathbf{V}(\boldsymbol{\theta}_0)}{n}\right)^{-1}{\mathbf{I}^{(1)}(\boldsymbol{\beta}_0)}(\frac{\boldsymbol{\delta}}{\sqrt{n}}+\boldsymbol{\beta}_0)$ is the non-centrality parameter and $p+1$ is the degree of freedom.
\end{theorem}

\begin{corollary}\label{corol:distribution of LLR}
    When $\boldsymbol{\beta}_0 = \boldsymbol{0}$ in Theorem~\ref{theo:distribution of LLR}, then under $H_0:\boldsymbol{\beta}=\boldsymbol{\beta}_0$,
    \begin{equation*}
        \tilde{\Lambda}_n {\stackrel{d}{\longrightarrow}} \chi^2_{p+1}\left(u_0\right),
    \end{equation*} where $u_0=\boldsymbol{0}$,
    whereas under $H_1: \boldsymbol{\beta}=\boldsymbol{\beta}_0+\frac{\boldsymbol{\delta}}{\sqrt{n}}$, 
    \begin{equation*}
        \tilde{\Lambda}_n {\stackrel{d}{\longrightarrow}} \chi^2_{p+1}\left(u_1\right),
    \end{equation*} where $u_1=\boldsymbol{\delta}^\top{\mathbf{I}^{(1)}(\boldsymbol{\beta}_0)}(\mathbf{V}(\boldsymbol{\theta}_0))^{-1}{\mathbf{I}^{(1)}(\boldsymbol{\beta}_0)}\boldsymbol{\delta}$.
\end{corollary}
\noindent Detailed proof of Theorem \ref{theo: distribution of average score}, \ref{theo:distribution of LLR} and Corollary \ref{corol:distribution of LLR} will be provided in the appendix. Intuitively, a large value of $\tilde{\Lambda}_n$ in \eqref{eq:llratio} indicates confidence in favor of $H_1$. Then, provided a significance level $\alpha$, the test will terminate and reject the null at the first time when $\tilde{\Lambda}_n \geq \chi^2_{p+1, 1-\alpha}\left(u_0\right)$, and it accepts the null hypothesis if no such time exists as soon as all collections of samples are obtained. Consequently, the pair $(T(\alpha), d(\alpha))$ of the decision rule in $\eqref{eq:HTE}$ is proposed in our setting as 
\begin{equation*}
        T(\alpha)=\inf \left\{n: \tilde{\Lambda}_n \geq \chi^2_{p+1, 1-\alpha}\left(u_0\right)\right\}, \quad d(\alpha)=1\left\{T(\alpha)<\infty\right\}.
\end{equation*}
Accordingly, we propose the always-valid $p$-value process $(p_n)^{\infty}_{n=1}$ at a sample size $n$ as
\begin{equation}\label{eq:p-value}
    p_n=\inf \{\alpha: T(\alpha) \leq n, d(\alpha)=1\}=\min _{k \leq n} \Pr(X \geq \tilde{\Lambda}_k), \quad\mbox{where}~X\sim \chi^2_{p+1}(u_0).
\end{equation}
It is obvious to see that $p_n$ is non-increasing in $n$ and, $p_{T(\alpha)}=\alpha$. Further, the power is asymptotically guaranteed to approach to one, by the following theorem.

\begin{theorem}\label{theo:test power}
For the test in \eqref{eq:HTEtest} and setting in Theorem \ref{theo:distribution of LLR},
\begin{equation*}
 \lim_{\delta \to \infty} \Pr(\tilde{\Lambda}_n \geq \chi^2_{p+1, 1-\alpha}\left(u_0\right)|\boldsymbol{\beta}=\boldsymbol{\beta}_0+\frac{\boldsymbol{\delta}}{\sqrt{n}}) = 1,
\end{equation*} where $\alpha$ is the significance level.
\end{theorem}
\noindent Detailed proof of Theorem \ref{theo:test power} is provided in the appendix. This allows for analysis of how the power of the test will change with $\boldsymbol{\delta}$ in the alternative and the sample size. The whole procedure of implementing the proposed POST method is wrapped up in Algorithm \ref{algor:Single POST}.
\begin{algorithm}
\caption{Single POST}
\label{algor:Single POST}
\begin{algorithmic}[1]
\STATE \textbf{Input: }$\boldsymbol{Y}, \mathbf{X}, \boldsymbol{\beta}_0, \alpha=0.05$
\STATE $\overline{\mathbf{S}}_n:=\frac{1}{n} \mathbf{S}_{n, \boldsymbol{\beta}}^{(1)}\left(\hat{\boldsymbol{\theta}}, \boldsymbol{\beta}_0\right)$
\STATE $\tilde{\Lambda}_n=2\log\frac{\psi_{\left(\overline{\mathbf{I}}_n^{(1)}\left(\boldsymbol{\beta_0}\right)\left(\boldsymbol{\beta}-\boldsymbol{\beta}_0\right), \frac{\overline{\mathbf{v}}_n\left(\hat{\boldsymbol{\theta}}\right)}{n}\right)}\left(\overline{\mathbf{S}}_n\right)}{\psi_{\left(\mathbf{0}, \frac{\overline{\mathbf{v}}_n\left(\hat{\boldsymbol{\theta}}\right)}{n}\right)}\left(\overline{\mathbf{S}}_n\right)}$
\STATE $p_n=P_{X \sim \chi^2_{p+1}(u_0)}(X \geq \tilde{\Lambda}_n)$
\IF {$p_n\leq\alpha$}
    \STATE $d=1$
\ELSE
    \STATE $d=0$
\ENDIF
\STATE \textbf{Output: } $d$
\end{algorithmic}
\end{algorithm}

\subsection{Multiple testing}
To accommodate more than one treatment variation over the baseline, the multiple testing problem needs to be dealt with in online A/B testing sometimes referred to as the so-called comparisons problem \cite{hsu1996multiple}. In such cases, challenges arise from the incident that the HTE may be reported as being statistically significant merely by chance. Accordingly, the proposed POST framework is further extended. Specially, in our sequential testing setting, the $p$-value proposed in \eqref{eq:p-value} retains its validity as its conventional definition in a fixed horizon. Additionally, when considering multiple testing, the family-wise error rate (FWER) and the false discovery rate (FDR) need to be controlled, where the FWER is the probability of making at least one false rejections and the FDR is the expected proportion of the false rejections. In our multiple testing scenario, three procedures are proposed to control the FWER and FDR, i.e. the Bonferroni Correction \cite{miller1981simultaneous}, Benjamini-Hochberg Procedure \cite{benjamini1995controlling} and Benjamini–Yekutieli Procedure \cite{benjamini2001control}. The proposed POST method can be readily demonstrated that applying three procedures to the set of sequential $p$-values ensures control over FWER or FDR. Officially, the sequential multiple comparisons under the POST framework are proposed as follows in Proposition~\ref{prop:Bonferroni Correction} to \ref{prop:Benjamini–Yekutieli}. 

\begin{proposition}[Bonferroni Correction (BC) for POST]\label{prop:Bonferroni Correction}
For an arbitrary stopping time $T$, compute the sequential p-values $\left(p_T^i\right)_{i=1}^m$ by \eqref{eq:p-value} for $m$ comparisons, and rearrange them in an increasing order, denoted as $p_T^{(1)}, \ldots, p_T^{(m)}$. Then reject the hypotheses $(1), \ldots, (j)$, where $j$ is the maximum such that $p_T^{(j)} \leq \frac{\alpha}{m}$.
\end{proposition}

\begin{proposition}[Benjamini-Hochberg (BH) Procedure for POST]\label{prop:Benjamini-Hochberg}
For an arbitrary stopping time $T$, compute the corresponding sequential p-values $\left(p_T^i\right)_{i=1}^m$ by \eqref{eq:p-value} for $m$ comparisons, and rearrange them in an increasing order, denoted as $p_T^{(1)}, \ldots, p_T^{(m)}$. Then reject the hypotheses $(1), \ldots, (j)$, where $j$ is the maximum such that $p_T^{(j)} \leq \frac{\alpha \cdot j}{m}$.
\end{proposition}

\begin{proposition}[Benjamini-Yekutieli (BY) Procedure for POST]\label{prop:Benjamini–Yekutieli}
For an arbitrary stopping time $T$, compute the corresponding sequential $p$-values $\left(p_T^i\right)_{i=1}^m$ by \eqref{eq:p-value} for $m$ comparisons, and rearrange them in an increasing order, denoted as $p_T^{(1)}, \ldots, p_T^{(m)}$. Then reject the hypotheses $(1), \ldots, (j)$, where $j$ is the maximum such that $p_T^{(j)} \leq \frac{\alpha \cdot j}{m \cdot \sum_{r=1}^m \frac{1}{r}}$.
\end{proposition}

In particular, the BC method shows excellent control over FWER, while it may be a bit conservative to detect true effects effectively. To overcome this limitation, the BH procedure is introduced, focusing on controlling the FDR. The BY procedure improves the BH procedure by avoiding correlated $p$-values, which is accounted by the term $\sum_{r=1}^m \frac{1}{r}$ \cite{benjamini2001control}. All the three techniques require a set of $p$-values for each comparison as an input and produce a collection of rejections. In practice, the BY procedure is encouraged and adopted in this article, taking into account correlations between $p$-values. The whole procedure of multiple testing using POST is wrapped up in Algorithm \ref{algor:Multiple POST using BY}, exemplified using the BY procedure, and it is easy to replace the BY procedure with the BC  method or the BH procedure.

\begin{algorithm}
\caption{Multiple POST using BY procedure}\label{algor:Multiple POST using BY}
\begin{algorithmic}[1]
\STATE \textbf{Input: }$\boldsymbol{Y}, \mathbf{X}, \boldsymbol{\beta}_0, \alpha=0.05$
\STATE $m=ncol(\boldsymbol{\beta}_0)$
\FOR{$1\leq i\leq m$}
    \STATE $\overline{\mathbf{S}}_{n,i}:=\frac{1}{n} \mathbf{S}_{n, \boldsymbol{\beta}}^{(1)}\left(\hat{\boldsymbol{\theta}}, \boldsymbol{\beta}_{0,i}\right)$
    \STATE $\tilde{\Lambda}_{n,i}=2\log\frac{\psi_{\left(\overline{\mathbf{I}}_n^{(1)}\left(\boldsymbol{\beta_{0,i}}\right)\left(\boldsymbol{\beta}-\boldsymbol{\beta}_{0,i}\right), \frac{\overline{\mathbf{v}}_n\left(\hat{\boldsymbol{\theta}}\right)}{n}\right)}\left(\overline{\mathbf{S}}_{n,i}\right)}{\psi_{\left(\mathbf{0}, \frac{\overline{\mathbf{v}}_n\left(\hat{\boldsymbol{\theta}}\right)}{n}\right)}\left(\overline{\mathbf{S}}_{n,i}\right)}$
    \STATE $p_n^i=P_{X \sim \chi^2_{p+1}(u_0)}(X \geq \tilde{\Lambda}_n,i)$
    \STATE $j=1$
    \IF {$p_n^i\leq \frac{\alpha i}{m \sum_{r=1}^m 1 / r}$}
        \STATE $j=\max\{i, j\}$
    \ENDIF
\ENDFOR
\FOR{$1\leq i\leq m$}
    \IF {$i\leq j$}
        \STATE $d^{(i)}=1$
    \ELSE
        \STATE $d^{(i)}=0$
    \ENDIF
\ENDFOR
\STATE \textbf{Output: } $\boldsymbol{d}$
\end{algorithmic}
\end{algorithm}

\section{Simulation}
\subsection{Simulation settings}
In this section, the performance of the proposed POST method is examined by identifying HTE with various data structures in both covariates and the response in generalized linear models using simulated data following the data generation mechanism \cite{yu2020new}. To be specific, a sequence of samples are generated for the covariates $\{X_1, \ldots, X_{p+1}\}$, where $p$ is set as 30, and $X_1$ is always set as 1 to represent an intercept. $\{X_2, \ldots, X_7\}$ are considered as two settings: (i) \textbf{NU} setting: they are i.i.d. generated from N$(0,1)$, N$(1,1)$, N$(2,1)$, U$(-1,1)$, U$(0,2)$ and U$(1,3)$, respectively; and (ii) \textbf{MVN} setting: they are generated from MVN$(\boldsymbol{0}, \boldsymbol{\Sigma^*})$, where $\boldsymbol{\Sigma^*}$ has diagonal elements of 1 and off-diagonal elements as 0.5, to allow for mild correlations between covariates. All the remaining covariates $\{X_{8}, \ldots, X_{31}\}$ are i.i.d. generated from the standard normal distribution. Further, to generate the response, we consider GLMs in \eqref{eq:HTE}. The true values of the nuisance parameter $\boldsymbol{\theta} = (\theta_1, \ldots, \theta_{p+1})^\top$ is set as $(0, 1,1,1,-1,-1,-1,0,\ldots,0)^\top$, indicating that the truly useful covariates are $\{X_2,\ldots, X_7\}$. $A_i$ is generated from Bernoulli distribution with a mean of 0.5. To introduce the HTE, for simplicity, the true values of $\boldsymbol{\beta}$ is set as a $(p+1)$-dimensional zero vector for the control group, and $(0, b, 0, 0, b, 0,\ldots, 0)^\top$ for the treatment group, so that the treatment effects are introduced by a normal covariate $X_2$ and a uniform one $X_5$, respectively. Three different link functions $g(\cdot)$ are considered, namely the identity link $g(z)=z$, the logit link $g(z)=\ln\frac{z}{1-z}$, and the log link $g(z)=\ln{z}$, so that the response $Y$ follows a normal, a Bernoulli and a Poisson distribution, respectively, and the effect size $b$ belongs to $\{0.1,0.15\}$ for identity, $\{0.5,0.75\}$ for logit, and $\{0.05,0.08\}$ for log links correspondingly. The motivation of using different values in $b$ for different link functions is to account for the magnitude of the response value. By such settings in the response and the HTE, the performance of the proposed POST can be supported with great confidence. The whole data generation combination is summarized in Table \ref{tab:generate_data}. 

\begin{table}
\footnotesize
\centering
\caption{Distributions for the covariates X and response Y, possible link functions $g(\cdot)$.}
\label{tab:generate_data}
\begin{tabular}{c|lll}
\hline
       & \multicolumn{3}{c}{Possible Distributions/Values}                                      \\ \hline
\multirow{3}{*}{X}         & \multicolumn{3}{l}{(i) \textbf{NU} : Mixture of N$(0,1)$, N$(1,1)$, N$(2,1)$, U$(-1,1)$, U$(0,2)$ and U$(1,3)$} \\
 & \multicolumn{3}{l}{(ii) \textbf{MVN}: \parbox[t]{0.6\linewidth}{MVN$(\boldsymbol{0}, \boldsymbol{\Sigma^*})$ where $\boldsymbol{\Sigma^*}$ has diagonal elements of 1 and}} \\ 
 & \multicolumn{3}{l}{$\quad \quad $ off-diagonal elements as 0.5}\\
 \hline
Y                          & \multicolumn{3}{l}{(i) Normal, (ii) Bernoulli, (iii) Poisson}      \\
\hline
$g(\cdot)$                 & \multicolumn{3}{l}{(i) linear: $g(z)=z$, (ii) logit: $g(z)=\log\frac{z}{1-z}$, (iii) log: $g(z)=\log z$}      \\ \hline
\multirow{2}{*}{$\boldsymbol{\beta}$} & \multicolumn{3}{l}{(i) treatment group: $(0,b,0,0,b,0,\dots,0)^{\top}_{31 \times 1}$}          \\
                           & \multicolumn{3}{l}{(ii) control group: $(0,\dots,0)^{\top}_{31 \times 1}$}                    \\ \hline
$\boldsymbol{\theta_0}$                   & \multicolumn{3}{l}{$(0, 1, 1, 1, -1, -1, -1, 0, \ldots, 0)^{\top}_{31 \times 1}$}                         \\ \hline
\multirow{6}{*}{b}         & \multirow{3}{*}{single testing}          & identity       & $\{0.1, 0.15\}$                \\
                           &                                          & logit          & $\{0.5, 0.75\}$                \\
                           &                                          & log            & $\{0.05, 0.08\}$               \\ \cline{2-4} 
                           & \multirow{3}{*}{multiple testing}        & identity       & $\{0.2, 0.4, 0.6, 0.8\}$       \\
                           &                                          & logit          & $\{1, 2, 3, 4\}$               \\
                           &                                          & log            & $\{0.1, 0.2, 0.3, 0.4\}$       \\ \hline
\end{tabular}
\end{table}

To examine HTE using the proposed POST method, two scenarios are addressed separately, i.e., a single A/B testing and multiple testing, respectively, compared with other methods, such as the SST method. For the single case, the hypothesis test of nullity on $\boldsymbol{\beta}$ is conducted, following the procedure in Algorithm \ref{algor:Single POST}, and the Type I error is employed to evaluate the test performance, which computes the rejection ratio among repeated experiments. For the multiple case, a group of $m=32$ hypotheses are examined by the proposed POST method using Algorithm \ref{algor:Multiple POST using BY}, where the null hypotheses are true for 24 of them, and the remaining 8 alternatives are true, with $b$ being equally distributed at $\{0.2,0.4,0.6,0.8\}$, $\{1,2,3,4\}$ and $\{0.1,0.2,0.3,0.4\}$ for the identity , logit and log link, respectively. The false discovery rate (FDR) and true positive rate (TPR) are employed, where the FDR is defined as the expected proportion of false rejections in null hypotheses
$$\text{FDR} = E\left[\frac{\text{false rejections in $H_0$}}{\text{true rejections in $H_0$} + \text{false rejections in $H_0$}}\right],
$$
and the TPR as that of correct rejections in true alternatives
\begin{equation*}
    \text{TPR} = E\left[\frac{\text{rejections in $H_1$}}{\text{rejections in $H_1$} + \text{acceptances in $H_1$}}\right].
\end{equation*}

To evaluate the selection accuracy of the proposed POST method, the coverage ratio and the filter ratio are employed, where the coverage ratio indicates the percentage that the coefficients of useful covariates are estimated to be non-zero, and the filter ratio indicates the percentage that the coefficients of useless covariates are estimated to be zero. The samples of data are generated in a sequential manner where each time a sequence of $n$ data points are generated for both treatment and control groups, up to a maximum size of $N$. The whole experiment will be terminated at the first stopping time if HTE is identified when $n<N$, or at the time when $n=N$. We set $n=100$ and $N=1000$. Each experiment is repeated 100 times.

\subsection{Simulation results}
As can be found in Table \ref{tab:single}, the proposed POST method successfully identifies the truly useful covariates indicated by the nuisance parameter $\boldsymbol{\theta}$ with both a great coverage ratio and filter ratio, while the SST method fails in the sense that the filter ratio is always 0 using the SST method in all three generalized linear models, which implies that variable selection is achieved in the POST method, as is expected. Further, the proposed POST method performs better in controlling the Type I error than the SST method in all three generalized linear models in Table \ref{tab:Type I error} and a larger valid detection power in Table \ref{tab:power}, since the SST method just displays a unreal power without controlling the Type I error except in the \textbf{NU} setting with the identity link. In particular, when the HTE truly exists under three GLMs, the POST with adaptive lasso penalty controls the Type I error better, and demonstrates higher detection power than using other penalties, such as MCP and SCAD. The rationale behind this lies in the precision of adaptive lasso in selecting true covariates compared to others, especially in log regression shown in Table \ref{tab:single} where only adaptive lasso penalty is capable of excluding covariates with zero coefficients while retaining those with non-zero coefficients. This precision contributes to greater efficiency in controlling Type I error and achieving excellent detection power simultaneously. Also, the detection power of the POST test tends to increase as the HTE size $b$ grows larger, approaching to one along with the collected batches of samples, which is consistent with Theorem~\ref{theo:test power}. Consequently, these findings underscore that the POST outperforms the SST in terms of accuracy and power, and the POST method with adaptive lasso surpasses MCP and SCAD. Detailed values of the estimated coeﬀicients are attached in the appendix.

\begin{table}[t]
\footnotesize
\centering
\caption{Coverage and filter ratios of the estimated coefficients under different combinations of $\boldsymbol{X}$ and Y. Note that the coverage ratio indicates the percentage that the coefficients of useful covariates are estimated to be non-zero, and the filter ratio indicates the percentage that the coefficients of useless covariates are estimated to be zero.}
\label{tab:single}
\begin{tabular}{cccccccccc}
\hline
\multirow{3}{*}{GLM}   & \multirow{3}{*}{Method} & \multicolumn{4}{c}{Coverage Ratio} & \multicolumn{4}{c}{Filter Ratio}   \\ \cline{3-10} 
 &      & \multicolumn{2}{c}{NU} & \multicolumn{2}{c}{MVN} & \multicolumn{2}{c}{NU} & \multicolumn{2}{c}{MVN} \\ \cline{3-10} 
&        & Mean  & Std   & Mean  & Std   & Mean  & Std   & Mean  & Std   \\ \hline
\multirow{5}{*}{Identity}   
& MCP      & >0.999 & <0.001     & >0.999 & <0.001     & 0.748 & 0.138 & 0.757 & 0.147 \\
& SCAD     & >0.999 & <0.001     & >0.999 & <0.001     & 0.765 & 0.124 & 0.699 & 0.147 \\
& AdaLasso & >0.999 & <0.001     & >0.999 & <0.001     & 0.665 & 0.129 & 0.456 & 0.131 \\
& MLE(SST method)      & >0.999 & <0.001     & >0.999 & <0.001     & <0.001 & <0.001     & <0.001 & <0.001     \\ \hline
\multirow{5}{*}{Logit} 
& MCP      & 0.894 & 0.065 & 0.892 & 0.054 & 0.287 & 0.138 & 0.495 & 0.131 \\
& SCAD     & 0.932 & 0.062 & 0.907 & 0.054 & 0.076 & 0.084 & 0.307 & 0.117 \\
& AdaLasso & 0.971 & 0.050 & 0.951 & 0.050 & 0.182 & 0.122 & 0.137 & 0.102 \\
& MLE(SST method)      & >0.999 & <0.001     & >0.999 & <0.001     & <0.001     & <0.001     & <0.001     & <0.001     \\ \hline
\multirow{5}{*}{Log} 
& MCP      & 0.976 & 0.083 & 0.977 & 0.049 & 0.602 & 0.242 & 0.699 & 0.195 \\
& SCAD     & 0.980 & 0.051 & 0.984 & 0.049 & 0.783 & 0.142 & 0.73  & 0.143 \\
& AdaLasso & >0.999 & <0.001     & >0.999 & <0.001     & 0.843 & 0.095 & 0.84  & 0.091 \\
& MLE(SST method)      & >0.999 & <0.001     & >0.999 & <0.001     & <0.001     & <0.001     & <0.001     & <0.001     \\ \hline
\end{tabular}
\end{table}
    
\begin{table}[t]
\footnotesize
\centering
\caption{Type I error for detecting HTE under different combinations of $\boldsymbol{X}$ and Y.}
\label{tab:Type I error}
\begin{tabular}{cccccccc}
\cline{1-7}
\multirow{3}{*}{GLM} &
\multirow{3}{*}{Method} &
\multirow{3}{*}{b} &
\multicolumn{2}{c}{NU} &
\multicolumn{2}{c}{MVN} & \\ \cline{4-7}
& & & \multicolumn{2}{c}{Type I error} &
      \multicolumn{2}{c}{Type I error} &
       \\ \cline{4-7}
& & & Mean  & Std  & Mean  & Std  & \\ \cline{1-7}
\multirow{5}{*}{Identity} 
& MCP       & 0 & 0.010               & 0.032            & 0.010               & 0.032            &  \\
& SCAD      & 0 & 0.010               & 0.032            & 0.010               & 0.032            &  \\
& AdaLasso & 0 & 0.010               & 0.032            & 0.030               & 0.048            &  \\
& MLE(SST method)       & 0 & 0.021               & 0.012            & 0.450               & 0.165            &  \\ \cline{1-7}
\multirow{5}{*}{Logit} 
& MCP       & 0 & 0.09                & 0.074            & 0.070               & 0.067            &  \\
& SCAD      & 0 & 0.070               & 0.068            & 0.050               & 0.071            &  \\
& AdaLasso & 0 & 0.050               & 0.071            & 0.050               & 0.071            &  \\
& MLE(SST method)       & 0 & \textgreater{}0.999 & \textless{}0.001 & \textgreater{}0.999 & \textless{}0.001 &  \\ \cline{1-7}
    \multirow{5}{*}{Log}    
& MCP       & 0 & 0.980               & 0.042            & 0.540               & 0.107            &  \\
& SCAD      & 0 & 0.630               & 0.263            & 0.390               & 0.160            &  \\
& AdaLasso & 0 & 0.010               & 0.032            & 0.010               & 0.032            &  \\
& MLE(SST method)       & 0 & \textgreater{}0.999 & \textless{}0.001 & \textgreater{}0.999 & \textless{}0.001 &  \\ \cline{1-7}
\end{tabular}
\end{table}
    
\begin{table}[t]
\footnotesize
\centering
\caption{Power of detecting HTE under different combinations of $\boldsymbol{X}$ and Y.}
\label{tab:power}
\begin{tabular}{ccccccc}
\hline
\multirow{3}{*}{GLM}                  & \multirow{3}{*}{Method} & \multirow{3}{*}{b} & \multicolumn{2}{c}{NU}    & \multicolumn{2}{c}{MVN}   \\ \cline{4-7} 
& & & \multicolumn{2}{c}{Power} & \multicolumn{2}{c}{Power} \\ \cline{4-7} 
& & & Mean & Std   & Mean   & Std    \\ \hline
\multirow{5}{*}{Identity}   
&           & 0.15 & 0.990  & 0.032  & 0.860  & 0.107  \\ 
& MCP       & 0.1  & 0.350  & 0.108  & 0.200  & 0.105  \\
&           & 0.15 & 0.980  & 0.042  & 0.840  & 0.084  \\
& SCAD      & 0.1  & 0.320  & 0.092  & 0.190  & 0.110  \\
&           & 0.15 & 0.980  & 0.042  & 0.810  & 0.088  \\
& AdaLasso & 0.1  & 0.380  & 0.103  & 0.290  & 0.129  \\
&           & 0.15 & >0.999 & <0.001 & 0.780  & 0.123  \\
& MLE(SST method)       & 0.1  & 0.880  & 0.079  & 0.880  & 0.103  \\
&           & 0.15 & >0.999 & <0.001 & 0.990  & 0.032  \\ \hline
\multirow{5}{*}{Logit} 
&           & 0.75 & >0.999 & <0.001 & >0.999 & <0.001 \\
& MCP       & 0.5  & 0.380  & 0.123  & 0.210  & 0.160  \\
&           & 0.75 & 0.880  & 0.103  & 0.660  & 0.217  \\
& SCAD      & 0.5  & 0.350  & 0.118  & 0.190  & 0.152  \\
&           & 0.75 & 0.870  & 0.095  & 0.670  & 0.164  \\
& AdaLasso & 0.5  & 0.410  & 0.152  & 0.320  & 0.103  \\
&           & 0.75 & 0.930  & 0.095  & 0.810  & 0.099  \\
& MLE(SST method)       & 0.5  & >0.999 & <0.001 & >0.999 & <0.001 \\
&           & 0.75 & >0.999 & <0.001 & >0.999 & <0.001 \\ \hline
\multirow{5}{*}{Log}      
& MCP       & 0.05 & >0.999 & <0.001 & 0.700  & 0.149  \\
&           & 0.08 & >0.999 & <0.001 & 0.860  & 0.126  \\
& SCAD      & 0.05 & 0.870  & 0.149  & 0.470  & 0.221  \\
&           & 0.08 & >0.999 & <0.001 & 0.800  & 0.133  \\
& AdaLasso & 0.05 & 0.640  & 0.126  & 0.420  & 0.123  \\
&           & 0.08 & >0.999 & <0.001 & 0.990  & 0.032  \\
& MLE(SST method)       & 0.05 & >0.999 & <0.001 & >0.999 & <0.001 \\
&           & 0.08 & >0.999 & <0.001 & >0.999 & <0.001 \\ \hline
\end{tabular}
\end{table}
    
\begin{figure}
\centering
\includegraphics[width=13cm]{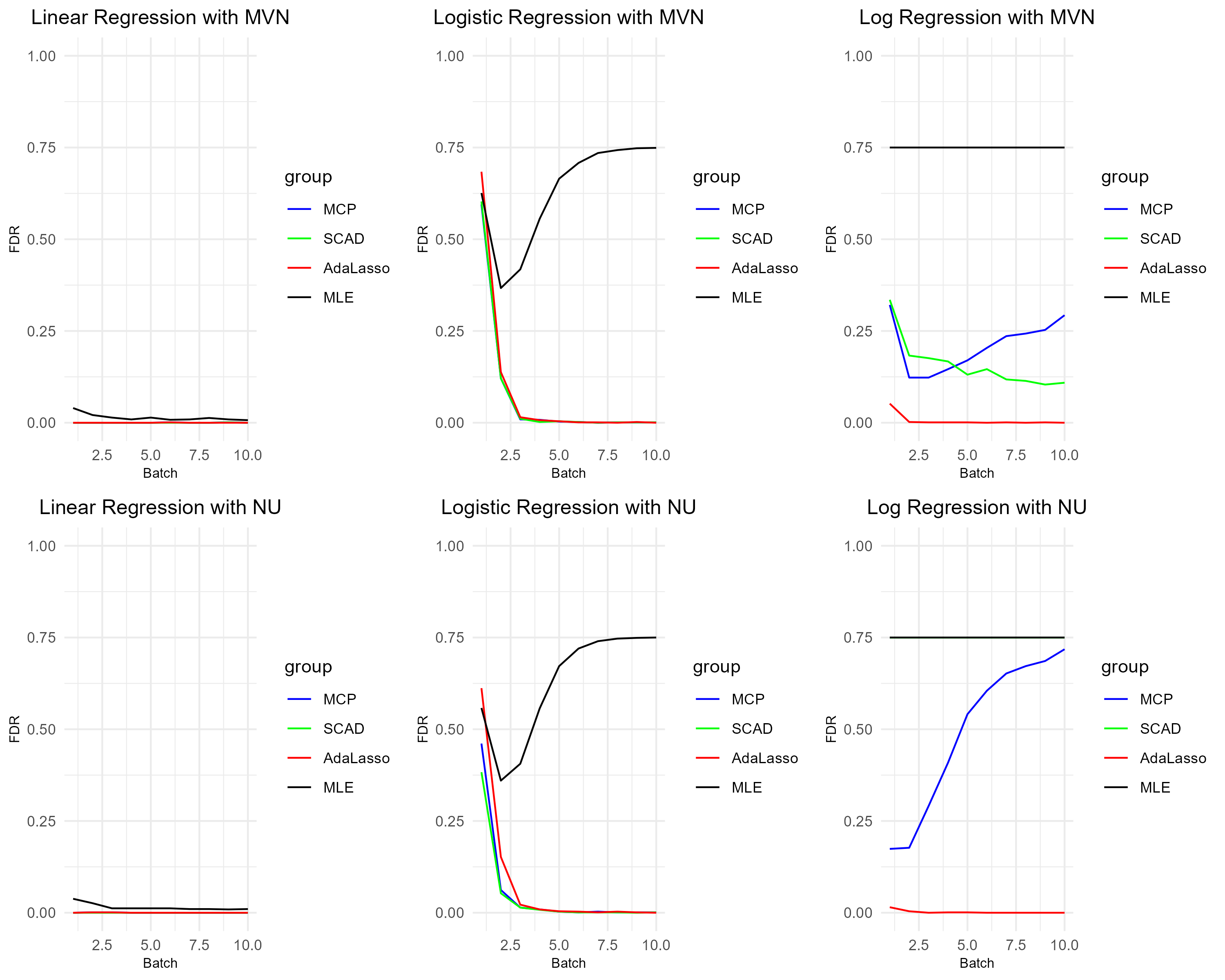}
\caption{Trends of FDR as batches increase in three regressions under two distributions of covariates.}
\label{fig:Line chart of FDR}
\end{figure}

\begin{figure}
\centering
\includegraphics[width=13cm]{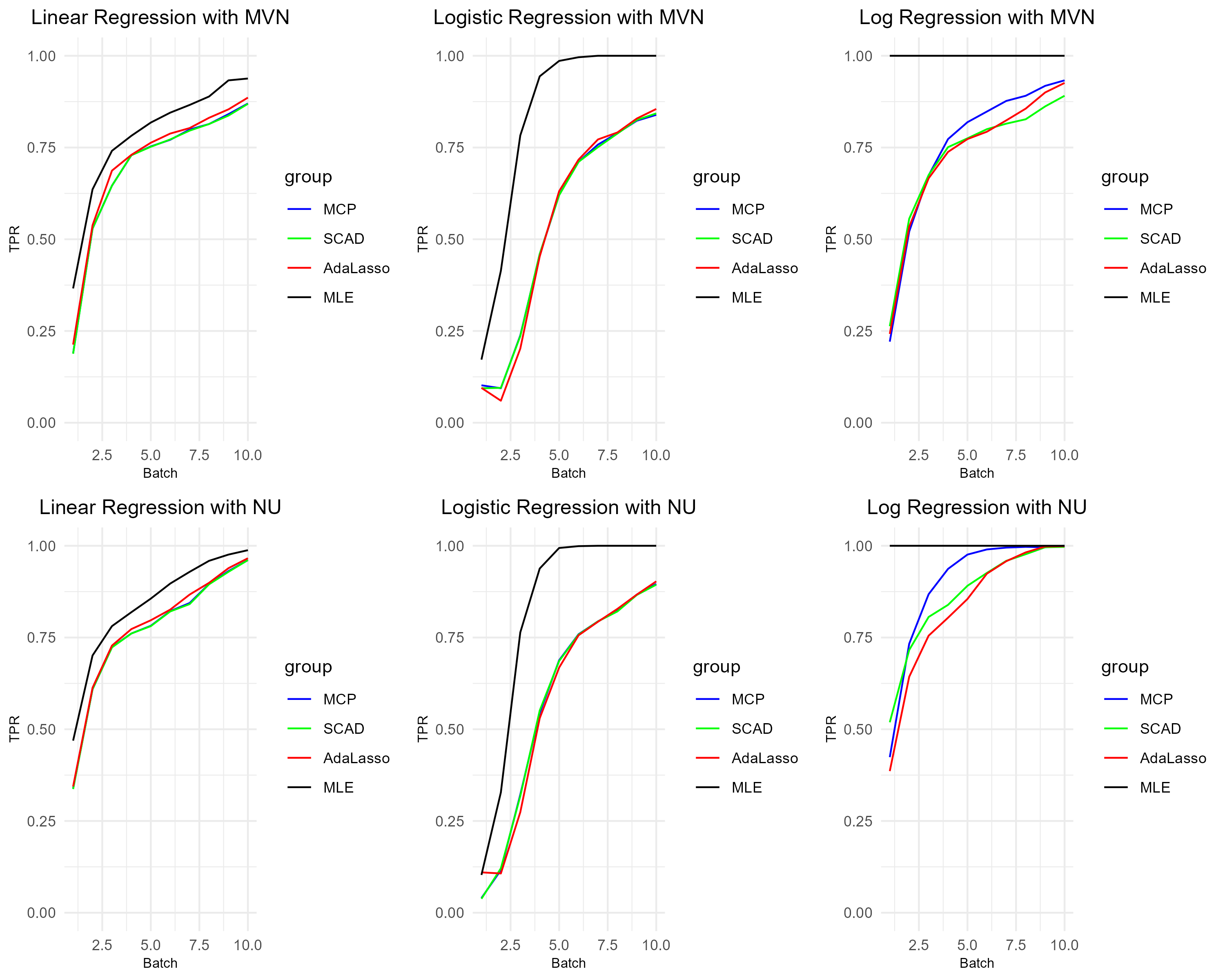}
\caption{Trends of TPR as batches increase in three regressions under two distributions of covariates.}
\label{fig:Line chart of TPR}
\end{figure}

In multiple testing, similar findings on the superb performance of the POST method are obtained in Table \ref{tab:FDR and TPR}. Specifically, the proposed POST method with adaptive lasso yields a high TPR and controls the FDR in three GLMs with covariates from different distributions, while the SST methods fails in most cases, even though it attains a relatively high but fake detection power. Additionally from Figure \ref{fig:Line chart of FDR} and \ref{fig:Line chart of TPR}, the trends of changes in FDR and TPR are visualized as batches increase, which clearly depict that the POST achieves a remarkable TPR approaching to one while concurrently minimizing FDR almost to zero, particularly with the adaptive lasso penalty. Clearly, the proposed POST method shows excellent performance in terms of the control over Type I error and detection power.
    
\begin{table}[t]
\footnotesize
\centering
\caption{FDR and TPR of detecting HTE for multiple testing under different combinations of $\boldsymbol{X}$ and Y.}
\label{tab:FDR and TPR}
\begin{tabular}{ccllllllll}
\hline
\multirow{3}{*}{GLM}   & \multirow{3}{*}{Method} & \multicolumn{4}{c}{NU}  & \multicolumn{4}{c}{MVN}   \\ \cline{3-10} 
&      & \multicolumn{2}{c}{FDR} & \multicolumn{2}{c}{TPR} & \multicolumn{2}{c}{FDR} & \multicolumn{2}{c}{TPR} \\ \cline{3-10} 
&      &\multicolumn{1}{c}{Mean} &
      \multicolumn{1}{c}{Std} &
      \multicolumn{1}{c}{Mean} &
      \multicolumn{1}{c}{Std} &
      \multicolumn{1}{c}{Mean} &
      \multicolumn{1}{c}{Std} &
      \multicolumn{1}{c}{Mean} &
      \multicolumn{1}{c}{Std} \\ \hline
\multirow{5}{*}{Identity}  
& MCP                     & <0.001     & <0.001     & 0.961      & 0.048      & <0.001     & <0.001     & 0.870      & 0.063      \\
& SCAD                    & <0.001     & <0.001     & 0.961      & 0.048      & <0.001     & <0.001     & 0.869      & 0.062      \\
& AdaLasso          & <0.001     & <0.001     & 0.966      & 0.046      & <0.001     & <0.001     & 0.886      & 0.055      \\
& MLE(SST method)                     & 0.010      & 0.025      & 0.988      & 0.027      & 0.007      & 0.020      & 0.938      & 0.059      \\ \hline
\multirow{5}{*}{Logit} 
& MCP                     & <0.001     & <0.001     & 0.896      & 0.056      & 0.001      & 0.007      & 0.839      & 0.061      \\
& SCAD                    & 0.001      & 0.007      & 0.894      & 0.059      & 0.001      & 0.007      & 0.843      & 0.064      \\
& AdaLasso          & <0.001     & <0.001     & 0.903      & 0.068      & <0.001     & <0.001     & 0.855      & 0.063      \\
& MLE(SST method)                     & 0.750      & 0.001      & >0.999     & <0.001     & 0.749      & 0.002      & >0.999     & <0.001     \\ \hline
\multirow{5}{*}{Log} 
& MCP                     & 0.718      & 0.116      & >0.999     & <0.001     & 0.293      & 0.316      & 0.933      & 0.092      \\
& SCAD                    & 0.266      & 0.312      & 0.997      & 0.016      & 0.109      & 0.201      & 0.891      & 0.075      \\
& AdaLasso          & <0.001     & <0.001     & 0.994      & 0.020      & <0.001     & <0.001     & 0.926      & 0.062      \\
& MLE(SST method)                     & 0.750      & <0.001      & >0.999     & <0.001     & 0.750      & <0.001      & >0.999     & <0.001     \\ \hline
\end{tabular}
\end{table}
    
\section{Real data analysis}
In this section, a comparative analysis is conducted to examine the performance of the proposed POST method using a real data set, the Diabetes Data from NIDDK (National Institute of Diabetes and Digestive and Kidney Diseases \footnote{https://www.niddk.nih.gov/}). The dataset contains instances extracted from a larger database, concentrating on females of Pima Indian heritage aged at least 21 years. For each instance, seven essential attributes are considered as covariates, including the number of pregnancies, plasma glucose concentration after a glucose tolerance test, diastolic blood pressure, triceps skin fold thickness, 2-hour serum insulin level (SIL), body mass index (BMI), diabetes pedigree function (DPF) and age, and the outcome of diabetes is a binary response to represent the presence (1) or absence (0) of diabetes. Specifically, the SIL is divided into two groups \cite{melmed2015williams} in a single test and the BMI is categorized into five groups in multiple tests \footnote{https://www.ncbi.nlm.nih.gov/books/NBK2003/}.

We examine the proposed POST method and compare with the SST method in both single and multiple testing scenarios. 
For single testing, our goal is to identify whether there is any serum insulin effect on diabetes outcomes of females, using the proposed POST method in Algorithm \ref{algor:Single POST} and the SST method, by firstly conducting an A/A test to show the validity of test on diabetes with the normal SIL and then an A/B test on events involving normal SIL or abnormal SIL, designating the former as the control group. 
SIL group consists of 496 instances for the normal group, and 272 for the abnormal one in A/B test, and to maintain a consistent sample size across both groups in the online test, 272 instances are randomly chosen for the normal SIL group as a truncation. Initiating the test with a sample size of 100 and subsequently increasing the sample size by 5 at a time for both groups, we compute the test statistics for POST and SST. As soon as the test statistic exceeds the predetermined critical value, the null hypothesis is rejected and the whole procedure is terminated. If all the data are used and no significant HTE is identified, we accept the null hypothesis. 
For multiple testing, five BMI categories are selected for pairwise comparisons, encompassing a total of $m=15$ comparisons, including 5 self-comparisons. Similarly, after sample truncation, each BMI group contains 98 instances. The always-valid $p$-value for each pair is calculated with the maximum sample size of $n=98$ from each BMI group using \eqref{eq:p-value}, and the proposed POST method is applied with the BY procedure using Algorithm \ref{algor:Multiple POST using BY}.

From Table \ref{tab:Fitted HTE}, it is found that the POST accepts the null hypothesis for A/A test while the SST method rejects at $n=350$, indicating that the Type I error is well controlled for the POST under the considered hypotheses but not for SST. For the A/B test, the POST and SST show similar performance and arrive at rejection of the null at the sample of 110 and 100 instances, respectively. This confirms that the POST method is able to detect the difference by accounting for the covariates with greater confidence. Further, based on the estimated HTE $\hat{\boldsymbol{\beta}}$ obtained from the logistic regression, SIL does show the HTE, passing the test at a significance level of $0.05$. In terms of multiple testing, 7 hypotheses out of 15 are rejected by the proposed POST method, and 8 by SST, and neither POST nor SST rejects the self-comparison experiments. The result indicates that there is indeed heterogeneous BMI effect on the outcome of diabetes across individuals.
    
\begin{table}[t]
\footnotesize
\centering
\caption{The fitted HTE and the $p$-values of covariates for Normal SIL and Abnormal SIL with the Diabetes data from NIDDK}
\label{tab:Fitted HTE}
\begin{tabular}{cccccccc}
\hline
\multirow{2}{*}{Covariates} & \multirow{2}{*}{Pregnancies} &\multirow{2}{*}{ Glucose} & {Blood} & {Skin} & \multirow{2}{*}{SIL}    & \multirow{2}{*}{BMI }    & \multirow{2}{*}{DPF}   \\ 
&&&Pressure&Thickness&&&\\
\hline
HTE       & 0           & -0.019  & -0.017        & -0.035        & 0.007  & -0.0005 & 0.613 \\
$p$-value    & 0.5         & 0.995   & 0.831         & 0.921         & 0.0424 & 0.507   & 0.132 \\ \hline
\end{tabular}
\end{table}

\section{Conclusions}
In this article, a novel online HTE testing method, the penalized online sequential test (POST), is proposed to detect the HTE for the generalized linear models in a high-dimensional setting. Firstly, the useful covariates are selected and estimated, and then the ratio of score functions under the null and alternative hypotheses is employed as the test statistic for POST. The asymptotic properties of the proposed test statistic are established. Furthermore, we develop an online $p$-value process for POST and expand the methodology to encompass online multiple testing. The validity of the proposed testing procedure is supported by theoretical derivations and empirical numerical analysis conducted on both simulated and real datasets, positioning our method as an effective tool for quick decision-making in high-dimensional online A/B testing scenarios.

\appendix

\section*{Appendix}

\section{Theoretical details}
    \begin{lemma}[Oracle property]\label{lemma:Oracle property}
        For an estimator $\hat{\boldsymbol{\theta}}=\left(\begin{array}{c} \hat{\boldsymbol{\theta}}^*_{\hat{\mathcal{A}}} \\ \boldsymbol{0} \end{array}\right)$ of $\boldsymbol{\theta}^*=\left(\begin{array}{c} \boldsymbol{\theta}^*_{\mathcal{A}} \\ \boldsymbol{0} \end{array}\right)$,
        where $\mathcal{A}=\left\{j: \theta_j^* \neq 0\right\}$, $|\mathcal{A}|=p_0$ , $\hat{\mathcal{A}}=\left\{j: \hat{\theta}_j^* \neq 0\right\}$, $|\hat{\mathcal{A}}|=\hat{p_0}$ and     $\boldsymbol{\Sigma^*}$ is the covariance matrix of $\boldsymbol{\theta}^*_{\mathcal{A}}$, then $\hat{\boldsymbol{\theta}}$ has oracle property if 
        
        1. Consistency in variable selection: 
        \begin{equation*}
            \lim _{n\to\infty }P\left(\hat{\mathcal{A}}=\mathcal{A}\right)=1.
        \end{equation*}
        
        2. Asymptotic normality: 
        \begin{equation*}
            \sqrt{n}\left(\hat{\boldsymbol{\theta}}^*_{\hat{\mathcal{A}}}-\boldsymbol{\theta}_{\mathcal{A}}^*\right) \stackrel{d}{\longrightarrow} \mathbf{M V N}_{p_0}\left(\mathbf{0}, \boldsymbol{\Sigma^*}\right).
        \end{equation*}
    \end{lemma}
    
    \paragraph{Theorem 1}\label{theo:distribution of average score in appn}
        For a generalized linear model in \eqref{eq:GLM} and a link function $\eta_i$ in \eqref{eq:HTE}, consider the following estimated average score $\overline{\mathbf{S}}_n$ for treatment group $(A_i=1)$ under $H_0: \boldsymbol{\beta}=\boldsymbol{\beta}_0$ :
    
        \begin{equation*}
            \overline{\mathbf{S}}_n:=\left.\frac{1}{n} \mathbf{S}_{n, \boldsymbol{\beta}}^{(1)}\left(\hat{\boldsymbol{\theta}}, \boldsymbol{\beta}\right)\right|_{\boldsymbol{\beta}=\boldsymbol{\beta}_0},
        \end{equation*}
        where $\hat{\boldsymbol{\theta}}$ is an estimator conform to oracle property of $\boldsymbol{\theta}$ calculated based on data from the control group $(A_i=0)$. Define the information matrix for treatment group $(A_i=1)$ as below:
        \begin{eqnarray*}
            \mathbf{I}^{(1)}(\boldsymbol{\beta}):=\mathbb{E}_{(\mathbf{X}, \mathbf{Y})}\left[\overline{\mathbf{I}}_n^{(1)}(\boldsymbol{\beta})\right]=\mathbb{E}_{(\mathbf{X}, \mathbf{Y})}\left[-\frac{1}{n} \frac{\partial \mathbf{S}_{n, \boldsymbol{\beta}}^{(1)}\left(\boldsymbol{\theta}, \boldsymbol{\beta}\right)}{\partial \boldsymbol{\beta}}\right], \quad and \\
            \mathbf{I}^{(1)}(\boldsymbol{\theta}):=\mathbb{E}_{(\mathbf{X}, \mathbf{Y})}\left[\overline{\mathbf{I}}_n^{(1)}(\boldsymbol{\theta})\right]=\left.\mathbb{E}_{(\mathbf{X}, \mathbf{Y})}\left[-\frac{1}{n} \frac{\partial \mathbf{S}_{n, \boldsymbol{\beta}}^{(1)}\left(\boldsymbol{\theta}, \boldsymbol{\beta}\right)}{\partial \boldsymbol{\theta}}\right]\right|_{\boldsymbol{\beta}=\boldsymbol{\beta}_0}.
        \end{eqnarray*}
        
        Then, under $H_0: \boldsymbol{\beta}=\boldsymbol{\beta}_0$,
        \begin{equation*}
            \sqrt{n} \overline{\mathbf{S}}_n {\stackrel{d}{\longrightarrow}} \mathbf{M V N}_{p+1}\left(\mathbf{0}, \mathbf{V}\left(\boldsymbol{\theta}_0\right)\right),
        \end{equation*}
        
        whereas under $H_1: \boldsymbol{\beta}=\boldsymbol{\beta}_0+\frac{\boldsymbol{\delta}}{\sqrt{n}}$, 
        \begin{equation*}
            \sqrt{n}\left(\overline{\mathbf{S}}_n-\mathbf{I}^{(1)}\left(\boldsymbol{\beta}_0\right)\left(\boldsymbol{\beta}-\boldsymbol{\beta}_0\right)\right) {\stackrel{d}{\longrightarrow}} \mathbf{M V N}_{p+1}\left(\mathbf{0}, \mathbf{V}\left(\boldsymbol{\theta}_0\right)\right),
        \end{equation*}
        where $\mathbf{V}(\boldsymbol{\theta})=\mathbf{I}^{(1)}(\boldsymbol{\theta})+\mathbf{I}^{(1)}(\boldsymbol{\theta})\boldsymbol{\Sigma}(\boldsymbol{\theta}) \mathbf{I}^{(1)}(\boldsymbol{\theta})$, $\boldsymbol{\Sigma}(\boldsymbol{\theta})$ is the covariance matrix of the penalized estimator of $\boldsymbol{\theta}$, and $\boldsymbol{\theta}_0$ is the true value of the nuisance parameter.
    
    \begin{proof}
        From now on, we assume $a_i(\phi)=a(\phi)=1$ for simplicity, and use $\mathbf{S}_n^{(1)}(\boldsymbol{\theta})$ in replace of $\mathbf{S}_{n, \boldsymbol{\beta}}^{(1)}\left(\boldsymbol{\theta}, \boldsymbol{\beta}_0\right)$ for ease of expression:
        \begin{equation*}
            \mathbf{S}_n^{(1)}(\boldsymbol{\theta})=\left.\sum_{i=1}^n\left(\frac{\partial \mu_i^{(1)}(\boldsymbol{\beta}, \boldsymbol{\theta})}{\partial \boldsymbol{\beta}^\top} \cdot \frac{\left(Y_i^{(1)}-\mu_i^{(1)}(\boldsymbol{\beta}, \boldsymbol{\theta})\right)}{V_i^{(1)}(\boldsymbol{\beta}, \boldsymbol{\theta})}\right)\right|_{\boldsymbol{\beta}=\boldsymbol{\beta}_0}.
        \end{equation*}
        
        From oracle property in Lemma~\ref{lemma:Oracle property}, we know that
        \begin{equation}\label{eq:original oracle}
            \sqrt{n}\left(\hat{\boldsymbol{\theta}}^*_{\hat{\mathcal{A}}}-\boldsymbol{\theta}_{\mathcal{A}}^*\right) \stackrel{d}{\longrightarrow} \mathbf{M V N}_{p_0}\left(\mathbf{0}, \boldsymbol{\Sigma^*}\right),
        \end{equation}
        where$\boldsymbol{\theta}_0=\left(\begin{array}{c} \hat{\boldsymbol{\theta}}^*_{\mathcal{A}} \\ \boldsymbol{0} \end{array}\right)$ is the true nuisance parameter. Let $\hat{\boldsymbol{\theta}}_0=\left(\begin{array}{c} \hat{\boldsymbol{\theta}}^*_{\hat{\mathcal{A}}} \\ \boldsymbol{0} \end{array}\right)$, then we transform \eqref{eq:original oracle} into \eqref{eq:oracle},
        \begin{equation}\label{eq:oracle}
            \sqrt{n}\left(\hat{\boldsymbol{\theta}}_0-\boldsymbol{\theta}_0\right) \stackrel{d}{\longrightarrow} \mathbf{M V N}_{p+1}\left(\mathbf{0}, \boldsymbol{\Sigma(\boldsymbol{\theta_0})}\right).
        \end{equation}
        
        Make a Taylor expansion of $\overline{\mathbf{S}}_n$ around $\boldsymbol{\theta}_0$ :
        \begin{eqnarray}\label{eq:Taylor expansion}
            \overline{\mathbf{S}}_n & = & \frac{1}{n} \mathbf{S}_n^{(1)}\left(\boldsymbol{\theta}_0\right)-\overline{\mathbf{I}}_n^{(1)}\left(\boldsymbol{\theta}_0\right)\left(\hat{\boldsymbol{\theta}}_0-\boldsymbol{\theta}_0\right)+\boldsymbol{O}_p\left(n^{-1}\right) \\
            \label{eq:Higher order infinitesimal}
            & = & \frac{1}{n} \mathbf{S}_n^{(1)}\left(\boldsymbol{\theta_0}\right)-\mathbf{I}^{(1)}\left(\boldsymbol{\theta}_0\right)\left(\hat{\boldsymbol{\theta}}_0-\boldsymbol{\theta_0}\right)+\boldsymbol{o}_p\left(n^{-\frac{1}{2}}\right).
        \end{eqnarray}
        Since the first two terms of \ref{eq:Higher order infinitesimal} come from treatment group ($A_i=1$) and control group ($A_i=0$) respectively, which means they are independent, we can derive their asymptotic distributions separately.By central limit theorem, under $H_0: \boldsymbol{\beta}=\boldsymbol{\beta}_0$, the standardized first term converges to the following distribution:
        \begin{equation}\label{eq:CLT}
            \frac{1}{\sqrt{n}} \mathbf{S}_n^{(1)}\left(\boldsymbol{\theta}_0\right) {\stackrel{d}{\longrightarrow}} \mathbf{M V N}_{p+1}\left(\mathbf{0}, \mathbf{I}^{(1)}\left(\boldsymbol{\theta}_0\right)\right).
        \end{equation}
        
        Combining \eqref{eq:oracle} and \eqref{eq:CLT} yields the asymptotic distribution of $\overline{\mathrm{S}}_n$ under $H_0$ :
        \begin{equation*}
            \sqrt{n} \overline{\mathbf{S}}_n {\stackrel{d}{\longrightarrow}} \mathbf{M V N}_{p+1}\left(\mathbf{0}, \mathbf{V}\left(\boldsymbol{\theta}_0\right)\right),
        \end{equation*}
        where
        $\mathbf{V}(\boldsymbol{\theta})=\mathbf{I}^{(1)}(\boldsymbol{\theta})+\mathbf{I}^{(1)}(\boldsymbol{\theta})\boldsymbol{\Sigma(\boldsymbol{\theta_0})} \mathbf{I}^{(1)}(\boldsymbol{\theta}).$
        
        However, the asymptotic distribution of $\mathbf{S}_n^{(1)}\left(\boldsymbol{\theta}_0\right)$ doesn't stay the same under $H_1: \boldsymbol{\beta}=\boldsymbol{\beta}_0+\frac{\boldsymbol{\delta}}{\sqrt{n}}$ . To derive its asymptotic distribution under $H_1$, it should be reformulated as:
        \begin{eqnarray*}
            \frac{1}{n} \mathbf{S}_n^{(1)}\left(\boldsymbol{\theta}_0\right) & = &\frac{1}{n} \mathbf{S}_n^{(1)}\left(\boldsymbol{\theta}_0, \boldsymbol{\beta}\right)+\left[\frac{1}{n} \mathbf{S}_n^{(1)}\left(\boldsymbol{\theta}_0\right)-\frac{1}{n} \mathbf{S}_n^{(1)}\left(\boldsymbol{\beta}, \boldsymbol{\theta}_0\right)\right] \nonumber\\
            & = &\frac{1}{n} \mathbf{S}_n^{(1)}\left(\boldsymbol{\theta}_0, \boldsymbol{\beta}\right)+\mathbf{I}^{(1)}\left(\boldsymbol{\beta}_0\right)\left(\boldsymbol{\beta}-\boldsymbol{\beta}_0\right)+\boldsymbol{o}_p\left(n^{-\frac{1}{2}}\right),
        \end{eqnarray*}
        where $\mathbf{S}_n^{(1)}(\boldsymbol{\theta}, \boldsymbol{\beta})=\sum_{i=1}^n\left(\frac{\partial \mu_i^{(1)}(\boldsymbol{\beta}, \boldsymbol{\theta})}{\partial \boldsymbol{\beta}^\top} \cdot \frac{\left(Y_i^{(1)}-\mu_i^{(1)}(\boldsymbol{\beta}, \boldsymbol{\theta})\right)}{V_1^{(1)}(\boldsymbol{\beta}, \boldsymbol{\theta})}\right)$. 
        
        The last equality follows similarly as from \eqref{eq:Taylor expansion} to \eqref{eq:Higher order infinitesimal} since $\boldsymbol{\beta}-\boldsymbol{\beta}_0=\boldsymbol{O}_p\left(n^{-\frac{1}{2}}\right)$.
        
        Thus, under $H_1: \boldsymbol{\beta}=\boldsymbol{\beta}_0+\frac{\boldsymbol{\delta}}{\sqrt{n}}$,\
        \begin{equation}\label{eq:distribution of score under H1}
            \sqrt{n}\left(\frac{1}{n} \mathbf{S}_n^{(1)}\left(\boldsymbol{\theta}_0\right)-\mathbf{I}^{(1)}\left(\boldsymbol{\beta}_0\right)\left(\boldsymbol{\beta}-\boldsymbol{\beta}_0\right)\right) {\stackrel{d}{\longrightarrow}} 
            \mathbf{M V N}_{p+1}\left(\mathbf{0}, \mathbf{I}^{(1)}\left(\boldsymbol{\theta_0}\right)\right).
        \end{equation}
        
        The final distribution \eqref{eq:distribution of average score under H1} can be obtained by combining (\ref{eq:oracle}), \eqref{eq:Taylor expansion} and \eqref{eq:distribution of score under H1} :
        \begin{equation}\label{eq:distribution of average score under H1}
            \sqrt{n}\left(\overline{\mathbf{S}}_n-\mathbf{I}^{(1)}\left(\boldsymbol{\beta}_0\right)\left(\boldsymbol{\beta}-\boldsymbol{\beta}_0\right)\right) {\stackrel{d}{\longrightarrow}} \mathbf{M V N}_{p+1}\left(\mathbf{0}, \mathbf{V}\left(\boldsymbol{\theta}_0\right)\right).
        \end{equation}
    \end{proof}
    
    \paragraph{Theorem 2}
        For a generalized linear model in $\eqref{eq:GLM}$, a link function $\eta_i$ in \eqref{eq:HTE}, $\tilde{\Lambda}_n$ in $\eqref{eq:llratio}$ and $\mathbf{I}^{(1)}(\boldsymbol{\beta})$, $\mathbf{I}^{(1)}(\boldsymbol{\theta})$ in Theorem 1, then under $H_0:\boldsymbol{\beta}=\boldsymbol{\beta}_0$,
            \begin{equation*}
                \tilde{\Lambda}_n {\stackrel{d}{\longrightarrow}} \chi^2_{p+1}\left(u_0\right),
            \end{equation*} where $u_0=-\boldsymbol{\beta}_0^\top{\mathbf{I}^{(1)}(\boldsymbol{\beta}_0)}\left(\frac{\mathbf{V}(\boldsymbol{\theta}_0)}{n}\right)^{-1}{\mathbf{I}^{(1)}(\boldsymbol{\beta}_0)}\boldsymbol{\beta}_0$ is the non-centrality parameter and $p+1$ is the degree of freedom,
            whereas under $H_1: \boldsymbol{\beta}=\boldsymbol{\beta}_0+\frac{\boldsymbol{\delta}}{\sqrt{n}}$, 
            \begin{equation*}
                \tilde{\Lambda}_n {\stackrel{d}{\longrightarrow}} \chi^2_{p+1}\left(u_1\right),
            \end{equation*} where $u_1=(\frac{\boldsymbol{\delta}}{\sqrt{n}}-\boldsymbol{\beta}_0)^\top{\mathbf{I}^{(1)}(\boldsymbol{\beta}_0)}\left(\frac{\mathbf{V}(\boldsymbol{\theta}_0)}{n}\right)^{-1}{\mathbf{I}^{(1)}(\boldsymbol{\beta}_0)}(\frac{\boldsymbol{\delta}}{\sqrt{n}}+\boldsymbol{\beta}_0)$ is the non-centrality parameter and $p+1$ is the degree of freedom.
    
    \begin{proof}
        Define the probability ratio
        \begin{equation*}
            \tilde{\lambda}_n=\frac{\underset{\beta \in H_1}{\max} \psi_{\left(\overline{\mathbf{I}}_n^{(1)}\left(\boldsymbol{\beta_0}\right)\cdot\left(\boldsymbol{\beta}-\boldsymbol{\beta}_0\right), \frac{\overline{\mathbf{v}}_n\left(\hat{\boldsymbol{\theta}}\right)}{n}\right)}\left(\overline{\mathbf{S}}_n\right)}{\psi_{\left(\mathbf{0}, \frac{\overline{\mathbf{v}}_n\left(\hat{\boldsymbol{\theta}}\right)}{n}\right)}\left(\overline{\mathbf{S}}_n\right)},
        \end{equation*} then we have

        \begin{align*}
            \tilde{\lambda}_n
            &= \underset{\beta \in H_1}{\max} \frac{1} {\sqrt{(2\pi)^{p+1}\left|\frac{\overline{\mathbf{v}}_n\left(\boldsymbol{\theta_0}\right)}{n}\right|}} \exp\left\{-\frac{1}{2}\left(\overline{\mathbf{S}}_n - \overline{\mathbf{I}}_n^{(1)}\left(\boldsymbol{\beta}_0\right)(\boldsymbol{\beta} - \boldsymbol{\beta}_0)\right)^\top \right. \nonumber \\
            &\qquad \left. \times \left(\frac{\mathbf{v}_n\left({\boldsymbol{\theta_0}}\right)}{n}\right)^{-1} \left(\overline{\mathbf{S}}_n - \overline{\mathbf{I}}_n^{(1)}\left(\boldsymbol{\beta}_0\right)(\boldsymbol{\beta} - \boldsymbol{\beta}_0)\right)\right\} \nonumber \\
            &\quad / \frac{1} {\sqrt{(2\pi)^{p+1}\left|\frac{\overline{\mathbf{v}}_n\left(\boldsymbol{\theta_0}\right)}{n}\right|}} \exp\left\{-\frac{1}{2}\overline{\mathbf{S}}_n \left(\frac{\overline{\mathbf{v}}_n\left({\boldsymbol{\theta_0}}\right)}{n}\right)^{-1} \overline{\mathbf{S}}_n\right\} \nonumber \\
            &= \underset{\beta \in H_1}{\max} \exp\left\{-\frac{1}{2}\left(-2\overline{\mathbf{S}}_n^\top \left(\frac{\overline{\mathbf{v}}_n\left({\boldsymbol{\theta_0}}\right)}{n}\right)^{-1} \overline{\mathbf{I}}^{(1)}\left(\boldsymbol{\beta}_0\right) (\boldsymbol{\beta} - \boldsymbol{\beta}_0) \right. \right. \nonumber \\
            &\qquad + (\boldsymbol{\beta} -\boldsymbol{\beta}_0)^\top\overline{\mathbf{I}}_n^{(1)}\left(\boldsymbol{\beta}_0\right) \left(\frac{\overline{\mathbf{v}}_n\left({\boldsymbol{\theta_0}}\right)}{n}\right)^{-1} \left. \left. \overline{\mathbf{I}}_n^{(1)}\left(\boldsymbol{\beta}_0\right)(\boldsymbol{\beta} - \boldsymbol{\beta}_0) \right)\right\}.
        \end{align*}
        from the definition of $\tilde{\Lambda}_n$,
        \begin{align} \label{eq:llratio in appn}
            \tilde{\Lambda}_n &= \max_{\beta\in H_1} \left\{\overline{\mathbf{S}}_n^\top \left(\frac{\overline{\mathbf{v}}_n\left({\boldsymbol{\theta_0}}\right)}{n}\right)^{-1} \overline{\mathbf{I}}^{(1)}\left(\boldsymbol{\beta}_0\right) (\boldsymbol{\beta} - \boldsymbol{\beta}_0) \right. \nonumber \\
            &\qquad - \left. (\boldsymbol{\beta} - \boldsymbol{\beta}_0)^\top\overline{\mathbf{I}}_n^{(1)}\left(\boldsymbol{\beta}_0\right) \left(\frac{\overline{\mathbf{v}}_n\left({\boldsymbol{\theta_0}}\right)}{n}\right)^{-1} \overline{\mathbf{I}}_n^{(1)}\left(\boldsymbol{\beta}_0\right)(\boldsymbol{\beta} - \boldsymbol{\beta}_0) \right\}.
        \end{align}
        Define 
        \begin{align*}
            f(\boldsymbol{\beta}) &= \overline{\mathbf{S}}_n^\top \left(\frac{\overline{\mathbf{v}}_n\left({\boldsymbol{\theta_0}}\right)}{n}\right)^{-1} \overline{\mathbf{I}}^{(1)}\left(\boldsymbol{\beta}_0\right) (\boldsymbol{\beta} - \boldsymbol{\beta}_0) \nonumber \\
            &\quad - (\boldsymbol{\beta} - \boldsymbol{\beta}_0)^\top\overline{\mathbf{I}}_n^{(1)}\left(\boldsymbol{\beta}_0\right) \left(\frac{\overline{\mathbf{v}}_n\left({\boldsymbol{\theta_0}}\right)}{n}\right)^{-1} \overline{\mathbf{I}}_n^{(1)}\left(\boldsymbol{\beta}_0\right)(\boldsymbol{\beta} - \boldsymbol{\beta}_0),
        \end{align*}
        then
        \begin{equation*}
            \frac{\partial f}{\partial \beta} =2\overline{\mathbf{I}}_n^{(1)}\left(\boldsymbol{\beta}_0\right) \left(\frac{\mathbf{v}_n\left({\boldsymbol{\theta_0}}\right)}{n}\right)^{-1}(\overline{\mathbf{I}}_n^{(1)}\left(\boldsymbol{\beta}_0\right)\boldsymbol{\beta}-\overline{\mathbf{S}}_n).
        \end{equation*}
        Since $f(\boldsymbol{\beta})$ is a quadratic form, its maximum value point is set where the first-order partial derivatives are all 0, and it is obvious that $f(\boldsymbol{\beta})$ obtains the maximum value at $\boldsymbol{\beta}=(\overline{\mathbf{I}}_n^{(1)}\left(\boldsymbol{\beta}_0\right))^{-1}\overline{\mathbf{S}}_n$.
        Substituting $\boldsymbol{\beta}=(\overline{\mathbf{I}}_n^{(1)}\left(\boldsymbol{\beta}_0\right))^{-1}\overline{\mathbf{S}}_n$ into \eqref{eq:llratio in appn}, we can get
        \begin{equation*}
            \tilde{\Lambda}_n = \overline{\mathbf{S}}_n^\top\left(\frac{\overline{\mathbf{v}}_n\left({\boldsymbol{\theta_0}}\right)}{n}\right)^{-1}\overline{\mathbf{S}}_n-\boldsymbol{\beta_0}^\top\overline{\mathbf{I}}_n^{(1)}\left(\boldsymbol{\beta}_0\right)\left(\frac{\overline{\mathbf{v}}_n\left({\boldsymbol{\theta_0}}\right)}{n}\right)^{-1}\overline{\mathbf{I}}_n^{(1)}\left(\boldsymbol{\beta}_0\right)\boldsymbol{\beta_0}.
        \end{equation*}
        Based on the asymptotic distribution of $\overline{\mathbf{S}}_n$, under null hypothesis $H_0: \boldsymbol{\beta}=\boldsymbol{\beta}_0$, 
        \begin{equation*}
            \overline{\mathbf{S}}_n^\top\left(\frac{\overline{\mathbf{v}}_n\left({\boldsymbol{\theta_0}}\right)}{n}\right)^{-1}\overline{\mathbf{S}}_n \sim \chi^2_{p+1},
        \end{equation*} whereas under local alternative $H_1: \boldsymbol{\beta}=\boldsymbol{\beta}_0+\frac{\boldsymbol{\delta}}{\sqrt{n}}$, 
        \begin{equation*}
            \overline{\mathbf{S}}_n^\top\left(\frac{\overline{\mathbf{v}}_n\left({\boldsymbol{\theta_0}}\right)}{n}\right)^{-1}\overline{\mathbf{S}}_n \sim \chi^2_{p+1}((\boldsymbol{\beta} - \boldsymbol{\beta}_0)^\top\mathbf{I}_n^{(1)}\left(\boldsymbol{\beta}_0\right) \left(\frac{\mathbf{v}_n\left({\boldsymbol{\theta_0}}\right)}{n}\right)^{-1} \mathbf{I}_n^{(1)}\left(\boldsymbol{\beta}_0\right)(\boldsymbol{\beta} - \boldsymbol{\beta}_0)).
        \end{equation*}
        Thus, then under $H_0:\boldsymbol{\beta}=\boldsymbol{\beta}_0$,
        \begin{equation*}
            \tilde{\Lambda}_n {\stackrel{d}{\longrightarrow}} \chi^2_{p+1}\left(u_0\right),
        \end{equation*} where $u_0=-\boldsymbol{\beta}_0^\top{\mathbf{I}^{(1)}(\boldsymbol{\beta}_0)}\left(\frac{\mathbf{V}(\boldsymbol{\theta}_0)}{n}\right)^{-1}{\mathbf{I}^{(1)}(\boldsymbol{\beta}_0)}\boldsymbol{\beta}_0$,
        whereas under $H_1: \boldsymbol{\beta}=\boldsymbol{\beta}_0+\frac{\boldsymbol{\delta}}{\sqrt{n}}$, 
        \begin{equation*}
            \tilde{\Lambda}_n {\stackrel{d}{\longrightarrow}} \chi^2_{p+1}\left(u_1\right),
        \end{equation*} where $u_1=(\frac{\boldsymbol{\delta}}{\sqrt{n}}-\boldsymbol{\beta}_0)^\top{\mathbf{I}^{(1)}(\boldsymbol{\beta}_0)}\left(\frac{\mathbf{V}(\boldsymbol{\theta}_0)}{n}\right)^{-1}{\mathbf{I}^{(1)}(\boldsymbol{\beta}_0)}(\frac{\boldsymbol{\delta}}{\sqrt{n}}+\boldsymbol{\beta}_0)$.
    \end{proof}
    
    \paragraph{Theorem 3}
        For test $H_0: \boldsymbol{\beta}=\boldsymbol{\beta}_0$  \quad \text{v.s.} \quad $H_1: \boldsymbol{\beta}=\boldsymbol{\beta}_0+\frac{\boldsymbol{\delta}}{\sqrt{n}} \quad(\boldsymbol{\delta} \neq 0)$, we have
        \begin{equation*}
            \lim_{\delta \to \infty} \Pr(\tilde{\Lambda}_n \geq \chi^2_{p+1, 1-\alpha}\left(u_0\right)|\boldsymbol{\beta}=\boldsymbol{\beta}_0+\frac{\boldsymbol{\delta}}{\sqrt{n}}) = 1.
        \end{equation*} where $\alpha$ is significance level for POST.
    
    \begin{proof}
        \begin{align*}
        & 
        \lim_{\delta \to \infty}  \Pr(\tilde{\Lambda}_n \geq \chi^2_{p+1, 1-\alpha}\left(u_0\right)|\boldsymbol{\beta}=\boldsymbol{\beta}_0+\frac{\boldsymbol{\delta}}{\sqrt{n}}) \\
        & = 
        \lim_{\delta \to \infty} \Pr\left(\chi^2_{p+1}\left(u_1\right) \geq \chi^2_{p+1, 1-\alpha}(u_0)\right) \\
        & = 
        \lim_{\delta \to \infty} \Pr\left(\chi^2_{p+1}\left((\frac{\boldsymbol{\delta}}{\sqrt{n}})^\top\left(\frac{\mathbf{V}(\boldsymbol{\theta}_0)}{n}\right)^{-1}\frac{\boldsymbol{\delta}}{\sqrt{n}}\right) \geq \chi^2_{p+1, 1-\alpha}\right) \\
        & = 
        \lim_{\delta \to \infty} \Pr\left(\chi^2_{p+1}\left({\boldsymbol{\delta}}^\top\left(\mathbf{V}(\boldsymbol{\theta}_0)\right)^{-1}\boldsymbol{\delta}\right) \geq \chi^2_{p+1, 1-\alpha}\right) \\
        & = 1
    \end{align*}
    \end{proof}

\section{Experimental details}
\begin{table}[t]
\footnotesize
\centering
\caption{Estimated coefficients for POST with identity link under different combinations of $\boldsymbol{X}$ and Y.}
\begin{tabular}{c|ccccccccc}
\hline
\multirow{3}{*}{GLM}                  & \multirow{3}{*}{Covariates} & \multicolumn{4}{c}{NU}                 & \multicolumn{4}{c}{MVN}                   \\ \cline{3-10} 
 &     & \multicolumn{4}{c}{Method}                   & \multicolumn{4}{c}{Method}                   \\ \cline{3-10} 
 &     & MCP    & SCAD   & AdaLasso & MLE    & MCP    & SCAD   & AdaLasso & MLE    \\ \hline
\multirow{31}{*}{Identity}   & V1                          & 0.061 & 0.061 & 0.000 & -0.052 & 0.029  & 0.029  & 0.000 & -0.003 \\
 & V2  & 1.041  & 1.041  & 0.977    & 0.980  & 1.050  & 1.050  & 0.976    & 0.978  \\
 & V3  & 1.005  & 1.005  & 0.993    & 1.008  & 0.997  & 0.997  & 0.946    & 0.978  \\
 & V4  & 0.992  & 0.992  & 1.012    & 1.005  & 1.012  & 1.012  & 0.958    & 1.032  \\
 & V5  & -0.977 & -0.977 & -0.977   & -1.015 & -1.007 & -1.007 & -1.002   & -0.983 \\
 & V6  & -1.047 & -1.047 & -0.963   & -1.001 & -1.009 & -1.009 & -0.959   & -1.003 \\
 & V7  & -0.988 & -0.988 & -1.023   & -0.955 & -0.984 & -0.984 & -0.973   & -1.021 \\
 & V8  & 0.000  & 0.000  & 0.000    & -0.009 & 0.000  & 0.000  & 0.000    & -0.025 \\
 & V9  & 0.000  & 0.000  & 0.000    & 0.032  & 0.000  & 0.000  & 0.000    & 0.018  \\
 & V10 & 0.000  & 0.000  & 0.000    & 0.006  & 0.000  & 0.000  & 0.000    & -0.023 \\
 & V11 & 0.000  & 0.000  & 0.000    & 0.004  & 0.000  & 0.000  & 0.000    & -0.014 \\
 & V12 & 0.000  & 0.000  & 0.000    & 0.012  & 0.000  & 0.000  & 0.000    & -0.022 \\
 & V13 & 0.000  & 0.000  & 0.000    & 0.009  & 0.000  & 0.000  & 0.000    & 0.013  \\
 & V14 & 0.000  & 0.000  & 0.000    & -0.010 & 0.000  & 0.000  & 0.000    & 0.031  \\
 & V15 & 0.000  & 0.000  & 0.000    & 0.037  & 0.000  & 0.000  & 0.000    & -0.030 \\
 & V16 & 0.000  & 0.000  & 0.000    & -0.001 & 0.000  & 0.000  & 0.000    & 0.018  \\
 & V17 & 0.000  & 0.000  & 0.000    & 0.011  & 0.000  & 0.000  & 0.000    & -0.020 \\
 & V18 & 0.000  & 0.000  & 0.000    & 0.005  & 0.000  & 0.000  & 0.000    & -0.038 \\
 & V19 & 0.000  & 0.000  & 0.000    & -0.026 & 0.000  & 0.000  & 0.000    & 0.021  \\
 & V20 & 0.000  & 0.000  & 0.000    & -0.009 & 0.000  & 0.000  & 0.000    & -0.026 \\
 & V21 & 0.000  & 0.000  & 0.000    & -0.008 & 0.000  & 0.000  & 0.000    & -0.003 \\
 & V22 & 0.000  & 0.000  & 0.000    & 0.012  & 0.000  & 0.000  & 0.000    & -0.054 \\
 & V23 & 0.000  & 0.000  & 0.000    & 0.007  & 0.000  & 0.000  & 0.000    & 0.011  \\
 & V24 & 0.000  & 0.000  & 0.000    & -0.022 & 0.000  & 0.000  & 0.000    & 0.052  \\
 & V25 & 0.000  & 0.000  & 0.000    & 0.022  & 0.000  & 0.000  & 0.000    & 0.042  \\
 & V26 & 0.000  & 0.000  & 0.000    & -0.020 & 0.000  & 0.000  & 0.000    & 0.024  \\
 & V27 & 0.000  & 0.000  & 0.000    & 0.005  & 0.000  & 0.000  & 0.000    & 0.024  \\
 & V28 & 0.000  & 0.000  & 0.000    & -0.020 & 0.000  & 0.000  & 0.000    & 0.025  \\
 & V29 & 0.000  & 0.000  & 0.000    & -0.008 & 0.000  & 0.000  & 0.000    & 0.001  \\
 & V30 & 0.000  & 0.000  & 0.000    & -0.009 & 0.000  & 0.000  & 0.000    & -0.031 \\
 & V31 & 0.000  & 0.000  & 0.000    & -0.030 & 0.000  & 0.000  & 0.000    & 0.029  \\ \hline
\end{tabular}
\end{table}

\begin{table}[t]
\footnotesize
\centering
\caption{Estimated coefficients for POST with logit link under different combinations of $\boldsymbol{X}$ and Y.}
\begin{tabular}{c|ccccccccc}
\hline
\multirow{3}{*}{GLM}                  & \multirow{3}{*}{Covariates} & \multicolumn{4}{c}{NU}                 & \multicolumn{4}{c}{MVN}                   \\ \cline{3-10} 
 &     & \multicolumn{4}{c}{Method}                   & \multicolumn{4}{c}{Method}                   \\ \cline{3-10} 
 &     & MCP    & SCAD   & AdaLasso & MLE    & MCP    & SCAD   & AdaLasso & MLE    \\ \hline
\multirow{31}{*}{Logit} & V1                          & 0.217 & 0.212 & 0.000 & -1.113 & -0.036 & -0.036 & 0.000 & 0.167  \\
 & V2  & 0.934  & 0.935  & 0.927    & 0.460  & 1.004  & 1.004  & 1.018    & 0.401  \\
 & V3  & 0.907  & 0.910  & 0.972    & 0.851  & 1.062  & 1.062  & 0.823    & 0.602  \\
 & V4  & 0.915  & 0.917  & 0.849    & 0.663  & 1.035  & 1.035  & 0.762    & 0.530  \\
 & V5  & -0.950 & -0.950 & -1.065   & -0.389 & -1.037 & -1.037 & -1.001   & -0.305 \\
 & V6  & -0.919 & -0.918 & -0.873   & -0.279 & -0.931 & -0.931 & -0.883   & -0.074 \\
 & V7  & -1.044 & -1.046 & -0.880   & -0.302 & -1.076 & -1.076 & -0.774   & -0.913 \\
 & V8  & 0.216  & 0.218  & 0.000    & -0.049 & 0.000  & 0.000  & 0.000    & -0.057 \\
 & V9  & 0.000  & 0.000  & 0.000    & -0.144 & 0.000  & 0.000  & 0.000    & 0.000  \\
 & V10 & 0.000  & 0.000  & 0.000    & 0.021  & 0.000  & 0.000  & 0.000    & -0.009 \\
 & V11 & 0.000  & 0.000  & 0.000    & 0.077  & 0.000  & 0.000  & 0.000    & 0.066  \\
 & V12 & -0.070 & -0.077 & 0.000    & 0.314  & 0.000  & 0.000  & 0.000    & 0.140  \\
 & V13 & 0.000  & -0.007 & 0.000    & -0.247 & 0.000  & 0.000  & 0.000    & -0.135 \\
 & V14 & 0.000  & 0.000  & 0.000    & -0.152 & 0.000  & 0.000  & 0.000    & -0.056 \\
 & V15 & 0.000  & 0.000  & 0.000    & 0.064  & 0.000  & 0.000  & 0.000    & 0.117  \\
 & V16 & 0.000  & 0.000  & 0.000    & 0.409  & 0.000  & 0.000  & 0.000    & 0.207  \\
 & V17 & 0.000  & 0.000  & 0.000    & -0.248 & 0.000  & 0.000  & 0.000    & -0.031 \\
 & V18 & 0.024  & 0.048  & 0.000    & -0.341 & 0.000  & 0.000  & 0.000    & 0.137  \\
 & V19 & 0.000  & 0.000  & 0.000    & 0.132  & 0.000  & 0.000  & -0.062   & 0.084  \\
 & V20 & 0.000  & 0.000  & 0.000    & -0.059 & 0.000  & 0.000  & 0.000    & -0.073 \\
 & V21 & 0.000  & 0.000  & 0.000    & -0.129 & 0.000  & 0.000  & 0.000    & 0.036  \\
 & V22 & 0.000  & -0.016 & 0.000    & 0.102  & 0.000  & 0.000  & 0.000    & 0.028  \\
 & V23 & 0.000  & 0.000  & 0.000    & 0.002  & 0.000  & 0.000  & 0.000    & -0.019 \\
 & V24 & -0.175 & -0.181 & 0.000    & 0.036  & 0.000  & 0.000  & 0.000    & -0.081 \\
 & V25 & 0.000  & 0.000  & 0.180    & -0.099 & 0.000  & 0.000  & 0.000    & 0.236  \\
 & V26 & 0.000  & 0.000  & 0.000    & 0.054  & 0.000  & 0.000  & 0.000    & 0.071  \\
 & V27 & 0.000  & 0.000  & 0.000    & -0.228 & 0.000  & 0.000  & 0.000    & -0.104 \\
 & V28 & 0.000  & 0.000  & 0.000    & 0.168  & 0.000  & 0.000  & 0.000    & -0.099 \\
 & V29 & 0.000  & 0.000  & 0.000    & -0.019 & 0.000  & 0.000  & 0.000    & -0.097 \\
 & V30 & 0.000  & 0.000  & 0.000    & 0.021  & 0.000  & 0.000  & 0.048    & -0.007 \\
 & V31 & 0.000  & 0.000  & 0.000    & 0.122  & 0.000  & 0.000  & 0.000    & -0.025 \\ \hline
\end{tabular}
\end{table}

\begin{table}[t]
\footnotesize
\centering
\caption{Estimated coefficients for POST with log link under different combinations of $\boldsymbol{X}$ and Y.}
\begin{tabular}{c|ccccccccc}
\hline
\multirow{3}{*}{GLM}                  & \multirow{3}{*}{Covariates} & \multicolumn{4}{c}{NU}                 & \multicolumn{4}{c}{MVN}                   \\ \cline{3-10} 
 &     & \multicolumn{4}{c}{Method}                   & \multicolumn{4}{c}{Method}                   \\ \cline{3-10} 
 &     & MCP    & SCAD   & AdaLasso & MLE    & MCP    & SCAD   & AdaLasso & MLE    \\ \hline
\multirow{31}{*}{Log}      & V1                          & 0.310 & 0.064 & 0.000 & 0.057  & 0.470  & 0.239  & 0.000 & 0.406  \\
 & V2  & 0.930  & 1.008  & 1.010    & 0.394  & 0.941  & 0.280  & 0.966    & 0.309  \\
 & V3  & 0.927  & 0.972  & 0.987    & 0.345  & 0.888  & 1.125  & 0.964    & 0.425  \\
 & V4  & 0.948  & 0.985  & 0.995    & 0.196  & 0.262  & 1.121  & 1.017    & 0.348  \\
 & V5  & -0.941 & -1.046 & -0.932   & -0.085 & -0.565 & -0.932 & -0.941   & -0.268 \\
 & V6  & -1.011 & -0.977 & -1.007   & -0.098 & -0.832 & -0.915 & -1.012   & -0.355 \\
 & V7  & -1.017 & -1.010 & -0.966   & 0.249  & -0.783 & -0.746 & -1.029   & -0.310 \\
 & V8  & 0.000  & 0.000  & 0.000    & -0.057 & 0.000  & 0.000  & 0.000    & -0.198 \\
 & V9  & 0.000  & 0.000  & 0.000    & 0.061  & 0.000  & 0.000  & 0.000    & 0.068  \\
 & V10 & 0.037  & 0.000  & 0.000    & -0.046 & 0.000  & 0.000  & 0.000    & -0.159 \\
 & V11 & 0.000  & 0.000  & 0.000    & 0.054  & 0.000  & 0.000  & 0.000    & -0.057 \\
 & V12 & 0.000  & 0.000  & 0.000    & 0.046  & 0.000  & 0.000  & 0.000    & -0.119 \\
 & V13 & 0.004  & 0.000  & 0.000    & -0.165 & 0.000  & 0.000  & 0.000    & 0.228  \\
 & V14 & -0.051 & 0.000  & 0.000    & -0.159 & 0.000  & 0.000  & 0.000    & -0.540 \\
 & V15 & 0.002  & 0.000  & 0.000    & 0.103  & 0.000  & -0.147 & 0.000    & 0.071  \\
 & V16 & 0.000  & 0.000  & 0.000    & -0.061 & 0.000  & 0.000  & 0.000    & 0.026  \\
 & V17 & 0.006  & 0.000  & 0.000    & -0.088 & 0.000  & 0.000  & 0.000    & 0.076  \\
 & V18 & 0.045  & 0.000  & 0.000    & -0.228 & 0.000  & 0.000  & 0.000    & -0.032 \\
 & V19 & 0.000  & 0.000  & 0.000    & 0.043  & 0.000  & 0.000  & 0.000    & -0.095 \\
 & V20 & 0.000  & 0.000  & 0.000    & -0.086 & 0.000  & 0.000  & 0.000    & 0.053  \\
 & V21 & 0.000  & 0.000  & 0.000    & -0.100 & 0.000  & 0.000  & 0.000    & 0.239  \\
 & V22 & 0.000  & 0.000  & 0.000    & 0.004  & 0.000  & 0.000  & 0.000    & -0.120 \\
 & V23 & 0.000  & 0.000  & 0.000    & 0.043  & 0.000  & 0.000  & 0.000    & 0.187  \\
 & V24 & -0.007 & 0.000  & 0.000    & -0.032 & 0.000  & 0.000  & 0.000    & -0.013 \\
 & V25 & 0.000  & 0.000  & 0.000    & -0.024 & 0.000  & 0.000  & 0.000    & -0.197 \\
 & V26 & 0.000  & 0.000  & 0.000    & -0.051 & 0.000  & 0.000  & 0.000    & -0.157 \\
 & V27 & 0.000  & 0.000  & 0.000    & -0.050 & 0.000  & 0.000  & 0.000    & 0.064  \\
 & V28 & -0.056 & 0.000  & 0.000    & -0.099 & 0.000  & 0.000  & 0.000    & 0.240  \\
 & V29 & 0.000  & 0.000  & 0.000    & 0.021  & 0.000  & 0.000  & 0.000    & 0.030  \\
 & V30 & 0.068  & 0.000  & 0.000    & -0.095 & 0.000  & 0.000  & 0.000    & -0.341 \\
 & V31 & -0.002 & 0.000  & 0.000    & 0.110  & 0.000  & 0.000  & 0.000    & -0.200 \\ \hline
\end{tabular}
\end{table}

\clearpage
\section*{Acknowledgments} 
The authors appreciate the reviewing team for their efforts to make the article better organized and presented. Dr. Xin Liu’s research is partially supported by National Natural Science Foundation of China [grant number 12201383], Shanghai Pujiang Program [grant number 21PJC056], and Innovative Research Team of Shanghai University of Finance and Economics [grant number 2020110930].

\bibliographystyle{chicago}
\bibliography{main}

\end{document}